\documentclass[fleqn,usenatbib]{mnras}

\usepackage{newtxtext,newtxmath}
\usepackage{amsmath}

\usepackage[T1]{fontenc}

\DeclareRobustCommand{\VAN}[3]{#2}
\let\VANthebibliography\thebibliography
\def\thebibliography{\DeclareRobustCommand{\VAN}[3]{##3}\VANthebibliography}

\usepackage{graphicx}	
\usepackage[usenames,dvipsnames]{color}

\usepackage{amsmath}	
\usepackage{adjustbox} 
\usepackage{natbib}
\usepackage{xcolor}

\title[Galactic feedback during cosmic reionisation]{SPICE: the connection between cosmic reionisation and stellar feedback in the first galaxies}

\author[A. Bhagwat et al.]
{Aniket Bhagwat$^1$\thanks{E-mail: abhagwat@mpa-garching.mpg.de},
Tiago Costa$^{1,2}$,
Benedetta Ciardi$^1$,
R{\"u}diger Pakmor$^1$ \& 
Enrico Garaldi$^1$
\\
$^1$Max Planck Institut f\"{u}r Astrophysik, Karl Schwarzschild Stra{\ss}e 1, D-85741 Garching, Germany\\
$^2$School of Mathematics, Statistics and Physics, Newcastle University, UK\\
}
\date{Accepted XXX. Received YYY; in original form ZZZ}

\pubyear{2015}

\begin{document}
\label{firstpage}
\pagerange{\pageref{firstpage}--\pageref{lastpage}}
\maketitle

\begin{abstract}
We present {\tt SPICE}, a new suite of radiation-hydrodynamic, cosmological simulations targeting the epoch of reionisation. The goal of these simulations is to systematically probe a variety of stellar feedback models, including "bursty" and "smooth" forms of supernova energy injection, as well as poorly-explored physical scenarios such as hypernova explosions and radiation pressure on dust. We show that even subtle differences in the behaviour of supernova feedback drive profound differences in reionisation histories, with burstier forms of feedback causing earlier reionisation. However, we also find that some global galaxy properties, such as the dust-attenuated luminosity functions and star formation main sequence, remain degenerate between models. In particular, we show that stellar feedback and its strength determine the morphological mix of galaxies emerging by $z \, = \, 5$ and that the reionisation history is inextricably connected to intrinsic properties such as galaxy kinematics and morphology. While star-forming, massive disks are prevalent if supernova feedback is "smooth", "bursty" feedback preferentially generates dispersion-dominated systems. Different modes of feedback produce different strengths of outflows, altering the interstellar/circumgalactic medium in different ways, and in turn strongly affecting the escape of Lyman continuum (LyC) photons. We establish a correlation between galaxy morphology and LyC escape fraction, revealing that dispersion-dominated systems have escape fractions 10-50 times higher than their rotation-dominated counterparts at all redshifts. At the same intrinsic luminosity, dispersion-dominated systems should thus preferentially generate large HII regions as compared to their rotation-dominated counterparts. Since dispersion-dominated systems are more prevalent if stellar feedback is more explosive, reionisation occurs earlier in our simulation with burstier feedback. We argue that statistical samples of post-reionisation galaxy morphologies (using both stellar and gaseous components) probed with telescopes such as JWST, ALMA and MUSE can constrain stellar feedback at $z > 5$ and models of cosmic reionisation. 
\end{abstract}




\section{Introduction}
\label{sec:introduction}
About $300,000 \, \rm yr$ after the Big Bang, protons and free electrons coalesce for the first time, leading to the formation of hydrogen and helium atoms. At this time, the Universe is mostly neutral. By $z \, = \, 5$ \citep[e.g.][]{2023Gaikwad}, however, the vast majority of hydrogen has transitioned to an ionised state. This $\sim 1 \, \rm Gyr$ period is known as the epoch of reionisation (EoR).
Reionisation is brought about by the ionising flux generated by stellar populations in the first galaxies \citep{1987Shapiro,Madau1999,Gnedin_2000,Robertson_2010,Robertson2015,Eide2020}, with quasars now thought to provide only a minor contribution \citep{Kulkarni2019AGN,Mason2019eps}. The ionising photon budget is set by the abundance of young stars, while the ability of photons to escape from galaxies is governed by the structure of the interstellar medium (ISM), which, in turn, is shaped by a multitude of `feedback' processes associated to star formation. Within the $\Lambda$CDM paradigm of galaxy formation, such processes must be invoked in order to prevent the overproduction of stars even at $z > 6$ \citep{hopkins2014galaxies,costa2014,somerville:15}.

These feedback processes include massive stellar winds \citep{1977Weaver,2015Mackey,Geen_2020,Lancaster_2021,Guszejnov_2022}, photo-ionisation and photo-heating \citep{geen2015detailed,2017Peters,KimKimOstrikerr2018II}, radiation pressure \citep{murray2005,2022Menon} and supernova explosions \citep{1986Dekel}.
The impact of stellar feedback on the phase structure of the ISM is well established  \citep{1977McKee,1985Chevalier,Walch2015SILCCI,Martizzi_2016,2017KimISMbubbles}.   
Feedback in the form of ionising radiation suppresses radiative cooling \citep{efstathiou1992,Gnedin_2000,Sommervilee2002,Okamoto_2008}.  
Strong radiation fields can also reduce the baryon fractions of galaxies via photo-heating,
further suppressing growth \citep{Wise_2008,Okamoto_2008,Hasegawa_2012,Wise_2012,2013Pawlik}. Supernova explosions heat the ISM \citep{1977McKee} and launch galactic outflows, expelling ISM material that would otherwise form stars \citep[e.g.][]{Dubois:08, Puchwein2013, Fielding_2018, Orr_2019, Martizzi_2019}.  
Theoretical studies have for decades emphasised the importance of turbulence in shaping the ISM. Turbulence can be driven by processes such as gravitational instability \citep{klessen2004gravoturbulent} and feedback from supernovae \citep{1981Larson,1987Solomon,2004Heyer,Federrath_2016}. Feedback-driven turbulence in the ISM can both inhibit and drive star formation by either providing support against gravitational collapse, or by creating strong density contrasts allowing for rapid collapse, hence strongly regulating star formation efficiency    \citep{krumholz2005general,2010Osteriker,Ostriker_2011,Faucher_Gigu_re_2013,SILCC32016}. 

By influencing the structure of the ISM, stellar feedback plays a direct role in cosmic reionisation. Recent studies highlight the importance of supernova-driven outflows in facilitating Lyman continuum (LyC, photons with energy $> 13.6$eV) escape \citep{Wise2009,kimm2014escape,trebitsch2017fluctuating, trebitsch2018escape, rosdahl2022} through the creation of low-density channels. 
The impact of stellar radiation itself on the ISM, however, remains less clear. 
While studies agree that photo-ionisation feedback suppresses star formation bursts by counteracting the formation of high-density peaks in the ISM \citep{rosdahl2015galaxiesthatshine,2017SILCC4,SILCCZOOM2018}, the role of radiation pressure is less settled. Analytical arguments \citep{murray2005, thompson2015} suggest that radiation pressure on dust should launch galactic winds if systems are sufficiently bright. Radiation-hydrodynamic (RHD) simulations, however, generally find that early stellar feedback (such as radiation pressure) acts to suppress outflows through a reduction in star formation and in supernova clustering \citep{kimm2018, costa2019, agertz2020, smith2021}. Recent observational evidence points to an `effective Eddington limit' in star-forming galaxies at $z > 6.5$, observed through an absence of systems with high star formation rates and high optical depths \citep{fiore2023}. 
Though possibly not a unique interpretation, such trends may suggest a link between dust obscuration and the strength of galactic outflows that is not captured by current models.

Besides questions surrounding the key feedback driving mechanisms, there are significant numerical uncertainties in the modelling of feedback in galaxy evolution simulations.
An ab initio treatment of supernova feedback remains challenging in cosmological simulations due to prohibitive resolution requirements. Cosmological simulations thus have to resort to `subgrid' models attempting to capture the impact of supernova feedback at the resolution scale, i.e. $\gtrsim 20 \, \rm pc$\footnote{In this work, we denote comoving coordinates as "cpc, ckpc" etc and physical coordinates as "pc, kpc" etc.}. 
While such subgrid models have become more sophisticated in the last decade \citep{rosdahl2017snap, hopkins2018model}, uncertainties persist. For instance, while commonly adopted strategies account for unresolved `PdV' work through a momentum boost \citep[e.g.][]{kimm2015towards}, current models may underproduce hot gas, essential for launching galactic outflows, if supernova explosions are not resolved \citep{Hu2019}. Even when enforcing the correct terminal momentum, many existing models have to boost supernova feedback in order to reproduce realistic galaxy star formation rates and masses. Missing physics, such as cosmic ray injection and transport \citep{2018Diesling}, have been proposed as possible ways to further strengthen stellar feedback \citep[e.g.][]{alvarez2022}.

Another possible approach to overcome limitations caused by insufficient resolution and decouple the impact of stellar radiation and supernova feedback from the host's ISM, is through the adoption of an effective model for galactic winds \citep{2003Springel, Pillepich_2017}, with a prescribed mass outflow rate and velocity. While this approach promotes good numerical convergence properties, it limits the simulations' insight into the detailed interaction between supernovae, stellar radiation and the ISM and properties of host galaxies, restricting its effects to the scale of the intergalactic medium (IGM). It also introduces difficulties in the generation of mock observables, such as Ly$\alpha$ or X-ray emission, to which the ISM contributes significantly.

The recent influx of observations from JWST has begun to provide statistical samples of galaxies at unprecedented resolution deep into reionisation. Probes such as spectral energy distribution (SED) fitting, emission line ratios and UV/H$\alpha$ SFR indicators help constrain star formation properties of high redshift galaxies. While a large number of studies find that bursty star formation \citep{atek2023jwst,dressler2023building,Tacchella_2023,langeroodi2023ultraviolet,endsley2023starforming,asada2023bursty} dominates the histories of $M_* \leq 10^9$~M$_\odot$ galaxies, there is also evidence of a smoother star formation channel or even a combination of the two \citep{ciesla2023identification,Dressler_2023}. 
Recently, a bursty star formation history (SFH) has been invoked to alleviate \citep{Shen2023UV,sun2023seen,sun2023bursty,steinhardt2023highestredshift} the so-called "too many too bright" galaxies problem \citep{Ferrara_2023,Boylan_Kolchin_2023}. Other studies find that galaxies with bursty SFH could be the main drivers of reionisation \citep{simmonds2023lowmass,endsley2023starforming,Endsley_2023,atek2023jwst}. Therefore, understanding the implications of different SFH in high redshifts galaxies is key, and careful theoretical modelling will help to interpret various observations. 
Theoretical studies such as \citet{Hartley_2016} and \citet{Furlanetto_2022} use semi-analytical models to understand the effect of a bursty SFH on reionisation, finding that the sizes of HII regions are strongly modulated by bursts. 

Over the last decade, RHD simulations of reionisation such as {\tt CROC} \citep{Gnedin_2014}, {\tt Renaissance} \citep{Oshea2015}, {\tt Aurora} \citep{Pawlik2017Aurora}, {\tt Technicolor Dawn} \citep{Finlator2018Tech},  
 {\tt SPHINX} \citep{SphinxRosdahl2018,Katz_2018,rosdahl2022}, {\tt CoDa} \citep{Ocvirk_2016,Ocvirk_2020,Lewis_2022} and {\tt THESAN} \citep{Kannan2022,Garaldi2022,Smith2022} have used variations of the stellar feedback prescriptions previously described to simulate the high redshift universe (see \citealt{gnedin2022modeling} for a detailed review). Different simulations adopt varied approaches toward their modelling, with simulations like {\tt THESAN}, {\tt CoDa} and {\tt CROC} focusing on the large scale reionisation process with fairly large volumes (though unable to resolve galaxy scale heights),  whereas {\tt SPHINX} focuses on resolving the internal structure of galaxies while compromising on volume. Due to the extreme cost of RHD simulations (tens of millions of cpu hours), most are performed for only a single fiducial model at the target resolution. The effect of variations of input baryonic physics has remained comparatively unexplored. 

Available RHD simulations provide either a statistical sample of galaxies modeled within the same feedback prescription, or a variety of models for single, zoom-in galaxies \citep{Pallottini_2023,Katz_2022}. A systematic and statistical study of the effect of different feedback models is thus missing.
The approach taken in this paper is to model galaxy formation and reionisation such that we maximize how well we resolve the multi-phase ISM (down to $\approx 28 \rm pc$  at $z=5$) in a cosmological volume ($L_{\rm box} \approx 14.8 \rm cMpc$). The emphasis of these simulations is on the variations in supernova feedback prescriptions in order to quantify the impact of numerical uncertainties and to uncover observationally-testable connections between stellar feedback and the properties of the galaxies that drive reionisation. We introduce {\tt SPICE}, a suite of simulations which include mechanical feedback from supernovae, stellar feedback in the form of ionising radiation (with on-the-fly radiation transport) and radiation pressure. Our simulations explore different models for supernova feedback with variations in explosion timing, supernova energies and the presence of hypernovae, predicted to exist in the pristine conditions of the early universe \citep{kobayashi2006galactic}. 
Within the {\tt SPICE} framework, we aim to understand how the effect of different modes of feedback manifests in observable properties of galaxies. Indeed, by varying only the feedback prescription while keeping everything else constant, we are able to quantify \emph{relative differences} between the feedback models.

The paper is organised as follows. In Section~\ref{sec:simulations} we describe the {\tt SPICE} suite of simulations setup along with the physics prescriptions that are included. In Section~\ref{sec:results} we examine our results on the effect of stellar feedback on star formation, reionisation, UV luminosity functions, galaxy morphologies and LyC escape fractions. In Sections~\ref{sec:discussion} and~\ref{sec:conclusion} we discuss and summarize our findings, respectively.

\section{Simulations}
\label{sec:simulations}
In this section, we describe the simulations performed and post-processed for this study. We perform a total of three flagship simulations to study the effect of variations in supernova feedback. Relevant global simulation parameters are summarized in Table~\ref{tab:OverviewSims}, while individual variations in feedback processes are briefly described in Table~\ref{tab:simulations}.  

All simulations are performed with {\tt RAMSES-RT} \citep{rosdahl2013ramses,rosdahl2015scheme}, which is a radiation-hydrodynamics extension of the Eulerian adaptive mesh refinement code {\tt RAMSES} \citep{teyssier2002cosmological}. {\tt RAMSES-RT} solves the coupled gas hydrodynamics, radiative transfer of stellar radiation along with the self-gravity and N-body dynamics of gas, dark matter and stars. {\tt RAMSES-RT} uses a first-order Godunov method with the M1 closure for the Eddington tensor \citep{levermore1984relating} to solve the radiation transport equation. We employ the 'Global–Lax–Friedrich' (GLF) Riemann solver to advect radiation between cells. 

The adaptive mesh nature of the code allows for the grid to be dynamically refined in order to obtain higher numerical resolution within sub-regions of the simulation domain. We summarize the models used for our simulations below, with sections \ref{sec:ICs}-\ref{sec:Halofinding} describing global setup within {\tt RAMSES-RT} and sections \ref{sec:SF}-\ref{sec:StellarFeedback} describing the novel combination of galaxy formation and feedback models we include in our simulations. 
\subsection{Initial conditions}
\label{sec:ICs}
The initial conditions (ICs) for the simulations are generated at $z=30$ using {\tt MUSIC2-monofonIC}\footnote{https://bitbucket.org/ohahn/monofonic/}~\citep{Hahn_2020,Michaux_2020} with the 2PPT/2LPT approach. A $\rm \Lambda CDM$ cosmological model is adopted, with parameters $\Omega_{\Lambda} = 0.6901$, $\Omega_{\rm m} = 0.3099$,  $\Omega_{\rm b} = 0.0489$, $H_0 = 67.74$ \ \rm km/s/Mpc, $n_{\rm s} = 0.9682$, $\sigma_8 = 0.8159$ \citep{2016A&A...594A..13P}. We assume primordial gas mass fraction 
$X=0.7545$ for Hydrogen (H) and $Y=0.2455$ for Helium (He), along with a metallicity floor (see~\ref{sec:thermochem}).
All simulation boxes have a length $L_{\rm box}=10$~cMpc/$h$, with $512^3$ dark matter particles, i.e. a mean mass $m_{\rm dm}=6.38 \times 10^5$~M$_\odot$. {\tt MUSIC2-monofonIC} treats dark matter and baryons within the two-fluid approximation, using the novel 2nd order propagator perturbation theory (2PPT) \citep{Uhlemann2019,Rampf_2020}. The 2PPT approach involves perturbing particle masses with a distribution centered at $m_{\rm dm}$ (see \citealt{Hahn_2020}) which suppresses discreteness errors.  
{\tt RAMSES-RT} evolves baryons from density and momentum fields which are discretized at fixed locations in Eulerian space. Traditionally, these Eulerian fields are generated from Lagrangian displacements and interpolated back onto Eulerian grids, thus introducing interpolation erros. The 2PPT approach directly yields these fields without ad-hoc interpolation \citep{2020Porqueres}, therefore providing accurate ICs for codes like {\tt RAMSES-RT}. We also use the "fixing" technique of \citet{Angulo_2016}, which fixes the amplitude of each density mode to the expectation of the input power spectrum, therefore ensuring that the ICs generated are representative of the average Universe. We have produced halo mass functions for our ICs and verified that they closely agree with expectations for an average region.

\begin{table}
    \begin{tabular}{c|c|c|}
    \hline
    Name & Value & Description \\ 
    \hline\hline 
    $L_{\rm box}$            & 10 [cMpc/h]& Box size \\
    $m_{\rm dm}$             & $6.38 \times 10^5$ [$\rm \rm M_\odot$]& Mean mass of dark matter particles \\
     $N_{\rm dm}$            & $512^3$ & Number of dark matter particles  \\
    $m_{\ast}$               & 975 [$\rm \rm M_\odot$] & Minimum mass of stellar particles \\
    $Z_{\rm floor}$          & $3.2 \times 10^{-4}$ $Z_\odot$  & Metallicity floor at initial redshift                 \\ 
    $f_{\rm esc}$            & 1.0         & Stellar birth cloud escape fraction \\
    \hline
    \end{tabular}
    \caption{Basic properties of the simulated cosmological boxes, including the variable name, its adopted value and a short description.}
    \label{tab:OverviewSims}
\end{table}

\begin{table*}
    \begin{tabular}{c|c|c|c|c|c|c|c|c|}
    \hline
    Name & $\rm t_{SNe}$ & $\rm E_{SN/HN}$ (erg) &  Description \\ 
    \hline\hline
    {\tt bursty-sn} & 10 Myr & 2$\times10^{51}$ & All SN in a single stellar particle explode at the same time and have the same energy \\
    {\tt smooth-sn} & 3-40 Myr & 2$\times10^{51}$ & The time delay of SN is stochastically sampled, but all SN have the same energy \\ 
    {\tt hyper-sn} & 3-40 Myr & $10^{50} - 2\times10^{51} (\rm SN) + 10^{52} (\rm HN)$ &  The time delay and energy of SN is stochastically sampled + metallicity-dependent HN rate  \\
    \hline
    \end{tabular}
    \caption{Description of various simulations and supernova feedback model variations. The columns list (from left to right) the name of the simulation, delay time for supernova events, energy per supernova event, brief description of the simulation.}
    \label{tab:simulations}
\end{table*}
\subsection{Gravity and hydrodynamics}
\label{sec:hydro}
The Euler hydrodynamic equations are solved employing a second order Godonov scheme based on a MUSCL-Hancock method. We use the "HLLC" Riemann solver \citep{toro1994restoration} to calculate fluxes across cell interfaces. Additionally, the MinMod slope limiter is used to construct gas variables at cell interfaces from the values at cell-centers. We adopt an adiabatic index of $\gamma$ = 5/3 (for an ideal monatomic gas) in order to close the relation between gas pressure and internal energy. The dynamics of collisionless dark matter and stellar particles are computed using the Poisson equation with a particle-mesh solver. The dark matter and stellar particles are projected onto a grid with a cloud-in-cell interpolation \citep{guillet2011simple}. We employ a multigrid solver \citep{guillet2011simple} to solve the Poisson equation up to a refinement level
of 14 ($\Delta x \approx 100 \rm pc$), while at more refined levels we adopt a conjugate gradient solver to improve the computational speed.
\subsection{Cooling and gas thermochemistry}
\label{sec:thermochem}
Gas cooling and heating is calculated as described in \cite{rosdahl2013ramses}\footnote{On-the-fly metal-line cooling is not coupled to local radiation, see \cite{katz2022ramses} for a self-consistent treatment.}. 
We evaluate the non-equilibrium ionisation states of H and He ({\textsc{HII} , \textsc{H}e\textsc{II} , \textsc{H}e\textsc{III}}) in each computational cell, and advect them with the gas, as passive scalars (details of the quasi-implicit method can be found in \citealt{rosdahl2013ramses}). They are fully coupled to the local ionising radiation field and for these primordial species, we calculate cooling and heating due to Bremsstrahlung, photoionisation, collisional ionisation, collisional excitation, Compton cooling off the cosmic microwave background, and di-electronic recombination. 

The cooling contribution from metals at $T > 10^4$~ K is computed using tables generated with CLOUDY (\citealt{ferland1998cloudy}, version 6.02), assuming photoionisation equilibrium with a \cite{haardt1995radiative} UV background. 
Instead, for $T \leq 10^4$~K  we adopt the fine structure cooling rates from \cite{rosen1995global}, which allow gas to cool radiatively down to $\approx300$K. We also assume a homogeneous initial gas metallicity floor of $Z = 3.2 \times 10^{-4}$~Z$_\odot$, which is used to mimic missing molecular hydrogen cooling channels and metal enrichment from Pop-III stars \citep{wise2011birth}, and allows the gas to cool below $T=10^{4}$~K. We also adopt the on-the-spot approximation, where emission of ionising photons from recombining gas is ignored, i.e. we assume that it is all absorbed locally, within the same cell.  

\subsection{Refinement strategy}
\label{sec:refine}

The adaptive mesh refinement nature of {\tt RAMSES-RT} allows for each parent gas cell to be split into 8 cells when certain conditions are satisfied (see below). The cell refinement level $\ell$ determines the width of each gas cell as $\Delta x_\ell \ = \ 0.5^{\ell} L_{\rm box}$. We allow refinement levels in the range $\ell=9-16$. At the coarsest level ($\ell_{\rm min}=9$) we have a minimum physical resolution of 4.8 kpc at $z=5$, while at the finest level ($\ell_{\rm max} = 16$) the maximum physical resolution is $\approx28$ pc at $z=5$. Cells start at the coarsest level and are adaptively refined to higher levels to increase the numerical resolution of the simulation. A cell is refined if the following criteria are met: 
\begin{enumerate}
    \item if $M_{\rm dm} + \frac{\Omega_{m}}{\Omega{b}} \ M_{\rm b} > 8\times m_{\rm dm}$ (quasi-Lagrangian refinement criterion), 
    where $M_{\rm dm}$ and $M_{\rm b}$ are the total dark matter and baryonic (i.e. gas + stars) mass within the cell, and $m_{\rm dm}$ is the mean dark matter particle mass.
    \item The cell width is larger than $\frac{1}{8}$ of the local Jeans length.
\end{enumerate}
We adopt a constant comoving resolution throughout time, meaning that we allow for cells to be refined to $\ell_{\rm max}$ at all redshifts. This allows for the maximum physical resolution to be the highest at very high redshifts ($\approx$10.72 pc at $z \, =\, 20$). 
\subsection{Halo and galaxy finding}
\label{sec:Halofinding}
To identify dark matter haloes and galaxies, we use the Adapta{\tt HOP} halo finder \citep{Aubert_2004,200Tweed} in the most massive sub-maxima mode. A triaxial ellipsoid is fit to each (sub-)halo and we check that if the virial theorem is satisfied within this ellipsoid, where the center is the location of the densest particle. In case this condition is not satisfied, the volume is iteratively decreased until an inner virialized region is obtained. The volume of this largest ellipsoidal virialized region is used to define the virial radius $R_{\rm vir}$ and mass $M_{\rm vir}$. For the halo finder, we require a minimum of 100 dark matter particles per halo. Halo finder parameters (as defined in Appendix B of \citealt{Aubert_2004}) used are: $N_{\rm SPH} = 32$, $N_{\rm HOP} = 16$, $\rho_{\rm th} = 80$, and $f_{\rm poisson} = 4$. Galaxies are also identified with Adapta{\tt HOP}. Here we require at least 100 star particles per galaxy. The galaxy-finder parameters are: $N_{\rm SPH} = 10$, $N_{\rm HOP} = 10$, $\rho_{\rm th} = 10^3$ , and $f_{\rm poisson}$ = 4.

\subsection{Star formation}
\label{sec:SF}
We employ the multi-freefall star formation model implemented by \cite{kretschmer2020forming}. This includes a model for subgrid turbulence (as described in section~\ref{sec:Turbulence}), yielding a variable star formation efficiency that depends on local conditions described in terms of the virial parameter and the turbulent Mach number (see below). In each gas cell individually where the gas density is above a threshold of $n_{\rm H} \geq 10 \rm cm^{-3}$, we adopt a standard Schmidt law, for which the star formation density is given as, 
\begin{equation}
    \centering
    \Dot{\rho_{*}} = \epsilon_{\rm ff} \frac{\rho}{t_{\rm ff}},
\end{equation}
where $\rho$ is the gas density, $t_{\rm ff} = \sqrt{3\pi/(32G\rho)}$ is the freefall time, and $\epsilon_{\rm ff}$ is the star formation efficiency per freefall time. Typically, in galaxy formation simulations the value of $\epsilon_{\rm ff}$ is assumed to be a constant in the range 1-3 percent \citep{agertz2011formation}, motivated by observations of inefficient star formation on galactic scales \citep{bigiel2008star}, as well as in Milky Way giant molecular clouds (GMCs) \citep{krumholz2007slow,murray2011star,lee2016observational,grisdale2019observed}. 
This model can produce local (i.e. within a single computational cell) $\epsilon_{\rm ff}$ values as low as 0.1$\%$ likely suppressed by feedback \citep{Ostriker_2011,Kimost_2015}, while also reaching values $>100\%$ (i.e. star formation faster than freefall time). Feedback modulates star formation in different parts of the galaxy such that the global (averaged over the entire galaxy) $\epsilon_{\rm ff}$ values produced are in agreement with observations (see Fig. 6 in \citealt{kretschmer2020forming}).  

We assume that the gas density within a supersonic turbulent medium such as the ISM is described by a log-normal probability distribution function (PDF) given as
\begin{equation}
    p(s) = \frac{1}{\sqrt{2\pi\sigma_{s}^2}}{\rm exp}\frac{(s-\overline{s})^2}{2\sigma_s^2},
\end{equation}
where $s = \rm ln(\rho/\overline{\rho})$, $\sigma_s$ is the variance of $s$, $ \overline{s} =-1/2\sigma_s^2$ is the mean logarithmic density, and $\overline{\rho}$ is the mean density of the cell.

The local star formation efficiency $\epsilon_{\rm ff}$, as computed in
\citet{hennebelle2011analytical, federrath2012star}, is
\begin{align}
    \epsilon_{\rm ff} &= \frac{\epsilon}{\phi} \int_{s_{\rm crit}}^{\infty} \frac{t_{\rm ff}(\overline{\rho})}{t_{\rm ff}(\rho)} \frac{\rho}{\overline{\rho}} p(s) ds \\
    &= \frac{\epsilon}{2\phi}{\rm exp}\left(\frac{3}{8}\sigma_s^2\right)  \left[ 1 \ + \ {\rm erf}\left(\frac{\sigma_s^2 - s_{\rm crit}}{\sqrt{2\sigma_s^{2}}}\right)  \right],
    \label{eq:sfeff}
\end{align}
where we assume that gas with density larger than a critical value, $s_{\rm crit}$, is converted into stars. The parameter $\phi=1/0.49$ takes into account the uncertainty in free-fall timescales for gas with different densities and $\epsilon=0.5$ accounts for the fact that not all gas above $s_{\rm crit}$ is converted into stars (both quantities are derived in \citealt{federrath2012star}). 
The critical density for star formation ($s_{\rm crit}$) is calculated using the model from \cite{krumholz2005general}. In order to extend the model to the subsonic regime, we allow star formation when a cell is gravitationally unstable ($\alpha_{\rm vir} < 1$) and $\mathcal{M} < 1$. The critical density accounting for both supersonic and subsonic regimes is
\begin{equation}
s_{\rm crit} =  {\rm ln} \left[\alpha_{\rm vir} \left(1 + \frac{2\mathcal{M}^4}{1 + \mathcal{M}^2}\right) \right]
\end{equation}
Here,
\begin{equation}
    \alpha_{\rm vir} = \frac{5 \sigma^2}{3G\rho\Delta x^2}
    \label{eq:alphavir}
\end{equation}
is the local virial parameter, which is an indicator of local stability. $\Delta x$ is the cell size, $G$ is the gravitational constant, $\sigma$ is the 1D turbulent velocity (see \ref{sec:Turbulence}) and $\mathcal{M} \ = \ \sigma/c_s$ is the local Mach number, where $c_s$ is the speed of sound.

This model therefore allows for two star formation channels. In the first channel, for which $\alpha_{\rm vir} < 1$ (independent of local $\mathcal{M}$), the entire computational cell is gravitationally unstable. In the second channel, collapse owing to large density fluctuations caused by supersonic turbulence, i.e. $\mathcal{M} \gg 1$ (see Figure 1 in \citealt{kretschmer2020forming}), becomes possible even if $\alpha_{\rm vir} > 1$. For reference,  typical conditions for star-forming regions within the Milky way are $\alpha_{\rm vir} \simeq 3-10$ with $\mathcal{M} \simeq 10-20$ \citep{Spilker_2022}, and correspond to $\epsilon_{\rm ff} \simeq 0.01-0.02$ \citep{murray2011star,grisdale2019observed}.

Using the local star formation efficiency and the gas mass within the cell, we compute the number of new stellar particles in a given timestep by sampling a Poisson distribution. The initial stellar mass of new stars is always an integer multiple of 970~M$_\odot$ ($\approx 95-97\%$ of stars in a simulation have a mass of 970~M$_\odot$), but it is capped such that we do not deplete more than 90\% of the mass of the star-forming cell. We also assume that each stellar population represents a fully sampled \cite{chabrier2003galactic} initial mass function (IMF). 
\subsection{Subgrid turbulence}
\label{sec:Turbulence}
Turbulence is modelled using a sub-grid scale approach \citep{schmidt2006localised}. In the implementation of \cite{kretschmer2020forming}, an additional equation for the turbulent kinetic energy is adopted to account for advection of turbulent energy and work done by turbulent pressure (\citealt{schmidt2014numerical} and \citealt{semenov2018galaxies}). This is given by
\begin{equation}
    \frac{\partial}{\partial t}K_{\rm T} + \frac{\partial}{\partial x_j}(K_{\rm T} \Tilde{v}_j) + P_{\rm T}\frac{\partial \Tilde{v}_j}{\partial x_j} = C_{\rm T} - D_{\rm T},
    \label{eq:turbKinetic}
\end{equation}
where the turbulent kinetic energy $K_{\rm T} = 3/2\rho\sigma^2_{\rm T}$, is related to the turbulent pressure by $P_{\rm T} = 2/3K_{\rm T}$, with $\sigma_{\rm T}$ representing the 1D turbulent velocity dispersion. The left-hand side of the equation includes terms describing the time evolution, advection and compression. The right-hand side contains creation and destruction terms ($C_T$ and $D_T$, respectively), which can be written as
\begin{equation}
    C_T = 2\mu_T\sum_{ij}\Bigg[\frac{1}{2}\Biggl(\frac{\partial \Tilde{v}_i}{\partial x_j} + \frac{\partial \Tilde{v}_j}{\partial x_i} \Biggl) - \frac{1}{3}\bigl(\nabla \cdot \Tilde{v}\bigl) \delta_{ij}\Bigg]^2  \hspace{0.15em} = \hspace{0.15em} \frac{1}{2}\mu_T\big|S_{ ij}\big|^2,
\end{equation}
and 
\begin{equation}
    D_T = \frac{K_T}{\tau_{\rm diss}},
\end{equation}
where $\big|S_{ij}\big|$ is the mean flow viscous stress tensor (this term is evaluated as in \citealt{schmidt2011fluid}), $\Tilde{v}$ is the mean flow variable (defined as $\Tilde{v} = \overline{\rho v}/\overline{\rho}$) and the sum is calculated over the nearest neighbours of the computational cell in question.
This model has two important parameters, the turbulent viscosity $\mu_T$ and the dissipation time-scale $\tau_{\rm diss}$ , which are related to the cells' size by 
\begin{equation}
        \mu_T = \overline{\rho} \Delta x \sigma\hspace{1em} {\rm and}  \hspace{1em} t_{\rm diss} = \frac{\Delta x}{\sigma}.
\end{equation}
Previous implementations of turbulent star formation sub-grid models consider an in-situ calculation \citep{kimm2017feedback} of the turbulent velocity dispersion \citep{Perret,trebitsch2017fluctuating,trebitsch2018escape,hopkins2018model}, which can be thought of as the stationary limit of the model used in this work (i.e. the creation and destruction terms are considered to be equal). In our case, the model accounts for the non-equilibrium dissipation of the turbulent kinetic energy using the density and velocity fields without modifying the hydrodynamic solver. Turbulent velocity dispersion is estimated by solving Eq.~\ref{eq:turbKinetic} such that $\sigma = \sqrt{2 K_{\rm T}/ \overline{\rho}}$. The resulting velocity dispersion is adopted in Eq.~\ref{eq:alphavir} to evaluate the Mach number and the virial parameter entering the star formation efficiency computation (Eq.~\ref{eq:sfeff}).  
\subsection{Supernova feedback}
\label{sec:SNeFeedback}

\begin{table*}
    \begin{tabular}{|c|c|c|c|c|c|c|c|c|}
    \hline
    Photon group & $\epsilon_0$ [eV] & $\epsilon_1$ [eV] & $\langle\epsilon\rangle$ [eV] & $\rm \sigma_{HI}$ [$\rm cm^{2}$] & $\rm \sigma_{HeI}$ [$\rm cm^{2}$] & $\rm \sigma_{HeII}$ [$\rm cm^{2}$] & $\rm \kappa_{abs}^0$ [$\rm g \ cm^{-2}$] & $\rm \kappa_{scat}^0$ [$\rm g \ cm^{-2}$] \\ 
    \hline\hline
    Infrared & 0.1 & 1 & 0.61 & 0 & 0 & 0 & 0 & 10 \\
    Optical & 1 & 13.6 & 5.51 & 0 & 0 & 0 & $10^3$ & 0\\
    UV I & 13.6 & 24.59 & 17.9 & $3.19 \times 10^{-18}$ & 0 & 0 & $10^3$ & 0 \\ 
    UV II & 24.59 & 54.42 & 32.32 & $6.27\times10^{-19}$ & $4.78\times10^{-18}$ & 0 & $10^3$ & 0\\ 
    UV III & 54.42 & $\infty$ & 67.77 & $7.37 \times 10^{-20}$ & $1.12 \times 10^{-18}$ & $8.71 \times 10^{-19}$ & $10^3$ & 0\\ 
    \hline \\
    \end{tabular}
    \caption{{\textit{Columns left to right:}} Photon groups, the lower and upper energies defining their energy interval, mean photon group energies
, ionisation cross-sections to HI, HeI and HeII, absorption and scattering opacities to dust for each photon group.}
    \label{tab:RTGroups}
\end{table*}
We adopt the mechanical feedback scheme introduced by \cite{kimm2014escape} and implemented as in \cite{kimm2015towards} (see \citealt{hopkins2014galaxies} for a similar setup). The key idea behind this model is to correctly capture the terminal momentum associated to the snowplow phase of a SN remnant, by injecting into the surrounding cells a radial momentum $p_{\rm SN}$ which depends on whether the adiabatic (Sedov-Taylor) phase is resolved. 
The model introduces a parameter $\chi$, which gives the ratio between the swept-up mass $M_{\rm swept}$ and the ejected mass $M_{\rm ej}$ for each neighbouring cell (number of neighbouring cells $N_{\rm nbor}$ = 48) as
\begin{equation}
     \chi \equiv dM_{\rm swept} / dM_{\rm ej}.
\end{equation}
This is compared to a threshold value of
\begin{equation}
    \chi_{\rm tr} \equiv  69.58 \ n_{\rm H}^{-4/17}E_{\rm 51}^{-2/17}Z^{\prime -0.28},
\end{equation}
where $E_{51}$ is the explosion energy of an individual SN in units of $10^{51}$ erg, $n_{\rm H}$ is the hydrogen number density in $\rm cm^{-3}$ and $Z^\prime  = {\rm{max}}[0.01,Z/{\rm Z}_\odot]$. Additionally,
\begin{equation}
    dM_{\rm ej} = (1 - \beta_{\rm sn}) M_{\rm ej} / N_{\rm nbor}, 
\end{equation}
and
\begin{equation}
    dM_{\rm swept} = \rho_{\rm nbor} \biggl(\frac{\Delta x}{2}\biggl)^2 + \frac{(1 - \beta_{\rm sn})\rho_{\rm host} \Delta x^3}{N_{\rm nbor}} + dM_{\rm ej}.
\end{equation} 
Here $\rho_{\rm host}$ is gas density of the SN host cell, and $\beta_{\rm sn}$ determines what fraction of the gas mass ($M_{\rm ej} \ + \ \rho_{\rm host}\Delta x^3$) is re-distributed to the host cell of a SN (see Fig. 15 of \citealt{kimm2014escape}). The parameter $\beta_{\rm sn}$ is set to 4/52 to distribute the mass as evenly as possible to the host and neighbouring cells when the cells are on the same level of refinement. 

If $\chi > \chi_{\rm tr}$ the adiabatic phase is not resolved and the momentum during the snowplow phase ($p_{\rm SN,snow}$) is injected to the neighbouring cells. Otherwise the momentum during the adiabatic phase ($p_{\rm SN,ad}$) is injected, i.e. 
\begin{equation}
p_{\rm SN} = \begin{cases}
    p_{\rm SN,ad} = \sqrt{2\chi M_{\rm ej}f_{\rm e}({\rm \chi}) E_{\rm SN}} & \hspace{2em} \text{$\chi < \chi_{\rm tr}$\, \rm ,} \\
    p_{\rm SN,snow} & \hspace{2em} \text{$\chi \geq \chi_{tr} \,,$}
\end{cases}
\end{equation}
where $f_{\rm e}(\rm \chi) \ = \ 1 - \frac{\chi - 1}{3(\chi_{\rm tr} - 1)}$ smoothly connects the two regimes. The input momentum during the snowplow phase is 
\begin{equation}
     p_{\rm{SN},\rm{snow}} = 3 \times 10^5 \ {\rm{km \ s^{-1} \rm M_\odot}} \ E^{16/17}_{51} \ n_{\rm{H}}^{-2/17} \ Z^{\prime-0.14},
\end{equation}
following \citet{blondin1998transition,thornton1998energy,geen2015detailed,kim2015momentum}.
\cite{geen2015detailed} have shown that the final radial momentum from SN in the snowplow phase can be augmented by photoionisation of their environments by massive stars. 
Following their approach, for each SN event, the local Str{\"o}mgen radius $r_{\rm S}$ is calculated and compared with the cell width $\Delta x$. If the Str{\"o}mgen radius is not well resolved ($\Delta x >> r_{\rm s}$), we adopt a final input momentum of 
\begin{equation}
     p_{\rm{SN,snow}} = p_{\rm SN, snow} {\rm exp}\biggl(-\frac{\Delta x}{r_{\rm s}}\biggl)\  + \ p_{\rm SN + PH} \biggl(1 - {\rm exp}\biggl[-\frac{\Delta x}{r_{\rm s}}\biggl]\biggl)
\end{equation}
where $p_{\rm SN+PH}$ is obtained using a fit to the results of \cite{geen2015detailed}.
 
Our assumption of a \cite{chabrier2003galactic} IMF implies that a fraction $\eta_{\rm sn} = 0.31$ of the initial mass of the stellar particle is recycled back into the ISM. 
Out of the recycled mass, a fraction of $0.05$ is in form of metals heavier than He \citep{chabrier2003galactic}. Due to computational cost, we do not store individual element abundances, but rather track the ejected metal mass as a metallicity per cell. 
Our choice of the IMF gives a fixed supernova rate of 0.01639 SN M$_\odot^{-1}$, which results in $15$ SN per star particle for our minimum stellar mass of 970 M$_\odot$.

\subsection{Supernova feedback variations}
\label{sec:feedbackvariations}
Given significant uncertainties in numerical models for supernova feedback (see Section~\ref{sec:introduction}), we explore a number of variations in the timing and energy injected per supernova event, keeping the basic implementation described in Section~\ref{sec:SNeFeedback} fixed. The aim of these model variations is to quantify how the behaviour of stellar feedback affects reionisation and galaxy properties at high redshift.

\subsubsection{Bursty supernova feedback} 
In our first model, labeled {\tt bursty-sn} due to the bursty star formation behaviour it drives, energy from all SN events associated to a given stellar particle is injected simultaneously. Thus, we assume that each stellar particle hosts one single SN explosion event at 10~Myr, i.e. equivalent to the mean time at which SN occur for a \cite{chabrier2003galactic} IMF.  
We adopt a progenitor mass of 19.1 M$_\odot$, and assume that each individual SN injects 2 × $10^{51}$ ergs into its neighbouring cells. Since large amounts of energy are deposited simultaneously at the same position, supernova feedback in {\tt bursty-sn} is particularly efficient. A similar feedback scheme has been previously adopted by the {\tt NewHorizonAGN} and {\tt Obelisk} simulations \citep{Dubois2021NewHorizon,trebitsch2021obelisk}.

\subsubsection{Smooth supernova feedback}
The second model we explore is referred to as {\tt smooth-sn}, because, unlike {\tt bursty-sn}, it produces a smooth star formation history. Previous studies \citep{kimm2015towards,su2018discrete,keller2022empirically,keller2022uncertainties} have shown that discretising SN injections in time is crucial. Indeed, for a given stellar population, SN with progenitor masses $>$ 20~M$_\odot$ can explode as early as 3 Myr, whereas SN with a progenitor mass of 8~M$_\odot$ can explode as late as 40 Myr. In {\tt smooth-sn}, the SN delay times are accounted for by randomly sampling the lifetimes of individual stars using the inverse sampling method of a polynomial fit for the integrated SN rate from {\tt STARBURST99} (see Fig 2. in \citealt{kimm2015towards}). 
As before, for each individual SN we assume an energy injection of $2\times10^{51}$ ergs and a progenitor mass of 19.1~M$_\odot$. This setup is a step toward a more physically-motivated feedback model as compared to {\tt bursty-sn}, and a variation has been previously adopted by the {\tt SPHINX} simulations \citep{SphinxRosdahl2018, 2022LyCSphinx}. 

\subsubsection{Variable supernovae and hypernovae feedback}

In this model, called {\tt hyper-sn}, we vary the injected supernova energy per explosion event and, in addition, include a contribution from hypernova events. Indeed, a (metallicity-dependent) fraction of type-II SN is theorized to be associated to very large energies \citep{kobayashi2006galactic}, in the range $10^{51} - 10^{53}$ erg. Such events are referred to as hypernovae (HN), and are expected to be more frequent in low metallicity environments  \citep{kitayama2005supernova,kobayashi2006galactic,kobayashi2007simulations,smidt2014population}.
Indeed, to explain chemical compositions observed in metal-poor stars, \citet{grimmett2020chemical} argue that the hypernova fraction (fraction of all type-II SN events) in the early universe needs to be $>50\%$, while it decreases with increasing metallicity to reach the estimated rate of $\sim 1\%$ in the local universe \citep{podsiadlowski2004rates}.
Simulations of star-clusters and idealized galaxies have tested the effect of hypernova on small scales \citep{su2018discrete,brown2022testing}, finding that HN feedback can quench dwarf galaxies for up to $\sim1$ Gyr, hence making them relevant to study during reionisation.  

In \texttt{hyper-sn}, we assign each stellar particle a metallicity dependant HN fraction $f_{\rm HN}$ given as
\begin{equation}
    f_{\rm HN} = {\rm max}\biggl[ 0.5 \ {\rm exp}\biggl(-\frac{Z_\ast}{0.001}\biggl), 0.01\biggl],
    \label{eq:HNrate}
\end{equation}
where $Z_\ast$ is the metallicity of the stellar particle. $f_{\rm HN}$ represents the fraction of all SN events a stellar particle undergoes that are classified as HN. We assume an initial HN fraction of $f_{\rm HN}=50\%$ to explore the extreme case of strong feedback in metal-free environments.
In case of an HN event, we inject an explosion energy of $10^{52}$ erg. A constant progenitor mass of 30~M$_\odot$ is assumed for all HN events \citep{grimmett2020chemical}. 

For SN explosions, we assume that progenitors with different masses (between 8-40 M$_\odot$, see \citealt{diaz2018progenitor}) explode at different times and inject different energies and masses into the surrounding medium. Thus, in addition to sampling the explosion times as described in the {\tt smooth-sn} model, we also stochastically assign different energies to each SN event following a normal distribution centered at $1.2\times10^{51}$ erg (best fit distribution using the Z9.6+W18 model as in \citealt{sukhbold2016core}), with minimum and maximum energies of $10^{50}$ and $2\times10^{51}$ ergs, respectively \citep{sukhbold2016core,diaz2018progenitor,diaz2021progenitor}.

\subsection{Radiative transfer}
\label{sec:RT}
{\tt{RAMSES-RT}} solves radiation transport by taking the first two moments of the radiative transfer equation and obtaining a system of conservation laws which is closed with the M1 closure of the Eddington tensor \citep{levermore1984relating}. Further details of the methods used for the injection, propagation, and interaction of the ionising radiation with hydrogen and helium are described in \cite{rosdahl2013ramses},
while the diffusion of multi-scattering IR radiation is followed with the trapped/streaming photon scheme presented in \cite{rosdahl2015scheme}.
Each photon group, defined by a frequency interval, is described in each grid cell by the radiation energy density and the bulk radiation flux, which corresponds approximately to the radiation intensity integrated over all solid angles. {\tt RAMSES-RT} solves the non-equilibrium evolution of the ionisation fractions of hydrogen and helium, along with photon fluxes and the gas temperature in each grid cell.

The RT time-step is subject to the Courant condition as radiation is advected on the AMR grid with an explicit solver. As the RT time-step can become prohibitively smaller than the hydrodynamic time-step, we subcycle the RT on each AMR level, and allow for a maximum of 500 RT steps performed after each hydro step (if the projected number of RT subcycles exceeds 500, the hydro time-step is decreased accordingly). During RT subcycles at each refinement level, radiation is not allowed to propagate between refinement levels and radiation fluxes are treated with the Dirichlet boundary conditions.
The subcycling scheme in the context of flux-limited diffusion is described in \cite{commercon2014fast}.
\begin{figure*}
    \centering
    \includegraphics[width=\textwidth]{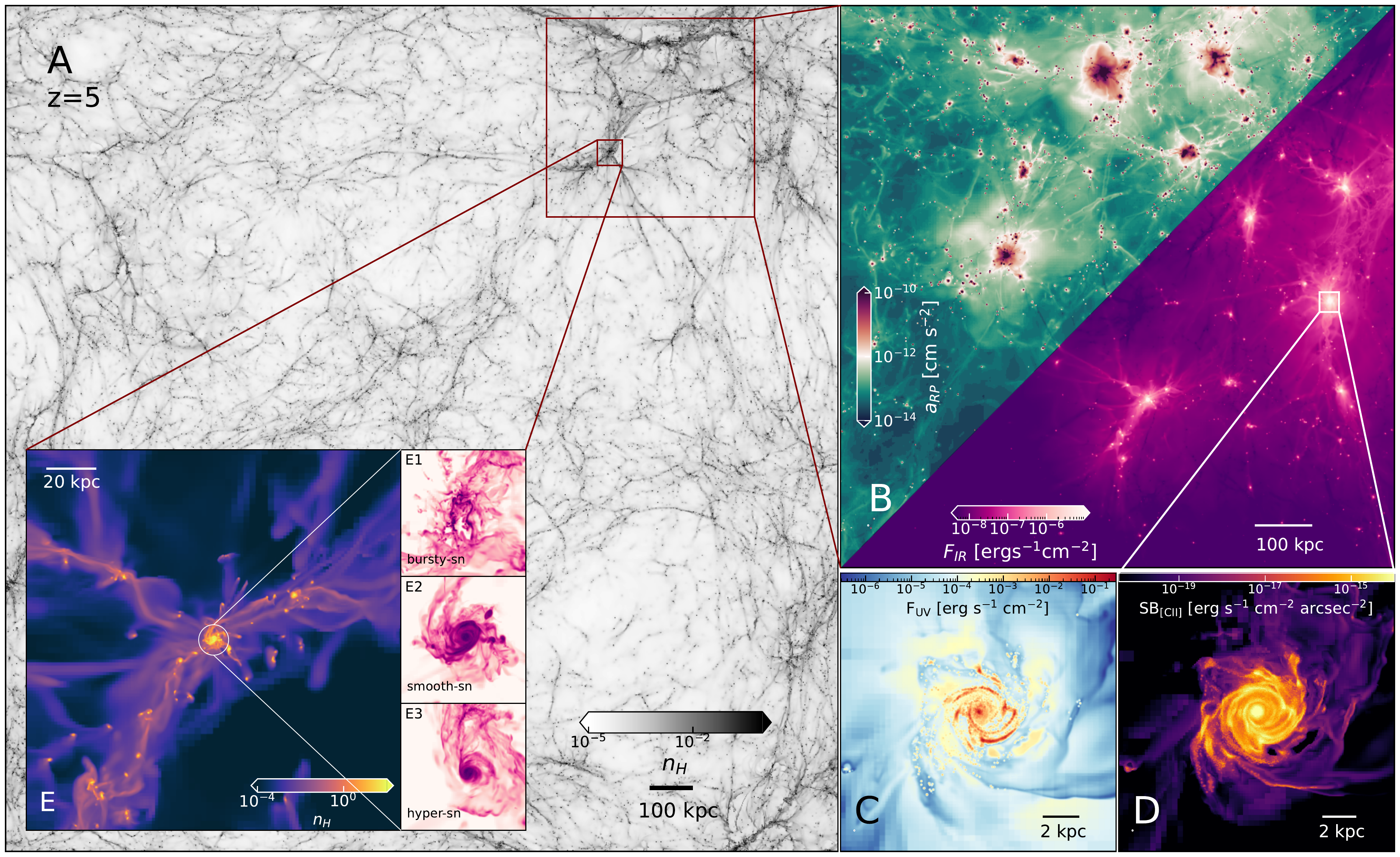}
    \caption{
    The {\tt SPICE} simulations and their scope and dynamic range. In panel {\tt \textbf{A}} we plot the volume-weighted gas density $z=5$, showing the cosmic web on the largest scales captured by the simulations. Panel {\tt \textbf{B}} shows a square region projected in two different quantities: the acceleration on gas produced by radiation pressure on dust (upper triangle), and infrared flux emergent from the galaxies (lower triangle). Panel {\tt \textbf{C}} shows the flux of ionising radiation escaping from a disc galaxy. We see channels of high (bright orange) LyC escape within the spiral arms of the galaxy and regions of inefficient escape (blue). {\tt SPICE} traces LyC radiation escape on scales of $\Delta x \approx 28 \rm pc$. In panel {\tt \textbf{D}} we show the surface brightness of the same galaxy in {\tt [CII]}. In panel {\tt \textbf{E}}, we zoom on a region of size $4\times R_{\rm vir}$ ($\approx 120 \rm kpc$) centred on a $\approx 2.7\times10^{11} \rm M_\odot$ halo. The central galaxy is fed by large-scale cold inflows. Sub-panels {\tt \textbf{E1,E2,E3}} highlight the very different gas density distributions emerging for this galaxy across our different simulations. In some simulations (e.g. \texttt{smooth-sn}, \texttt{hyper-sn}), gas has settled into a disc, while in others (\texttt{bursty-sn}), the gas component is strongly disrupted by strong outflows. Note that quantities shown in panels~{\tt \textbf{A,B,C,D} and \textbf{E}} are taken from {\tt smooth-sn} and can vary between models.
    }
    \label{fig:miniGigaplot}
\end{figure*}

To prevent prohibitively small time-steps and a large number of RT subcycles, we adopt the 'reduced speed of light approximation' (RSLA) \citep{Gnedin_2001} with the global reduced speed of light set to $\Tilde{c} = 0.1c$. Note that such approximation is used to advect the radiation field, while processes such as radiation pressure are treated with the full speed of light \citep{rosdahl2015scheme}. Studies performed by \cite{Costa2018a,Costa2018b} demonstrate a mild effect of the RSLA on the spatial extend of outflows driven by radiation pressure. When adopted in cosmological simulations targeting reionisation \citep{Gnedin_2016,Ocvirk_2016,2019Deparis}, the RSLA is found to affect mainly the post-reionisation neutral hydrogen fractions, which tend to be overestimated for a lower speed of light, while the differences are minimal during the overlap phase. Previous large scale simulations of reionisation \citep{SphinxRosdahl2018,Ocvirk_2020,Kannan_2021,2022LyCSphinx} adopt different values for $\Tilde{c}$, in the range $(0.01-0.2)c$, while finding convergence with values as low as $\Tilde{c} = 0.1c$ \citep{2019WuLya}. Therefore, we use results from these studies and adopt $\Tilde{c} = 0.1c$. 
\subsection{Stellar radiative feedback}
\label{sec:StellarFeedback}
We sample radiation fluxes from stellar populations in five photon frequency groups, i.e. infrared, optical and three bands of ionising ultraviolet photons, whose widths are determined by the ionising potentials of {\textsc{HI}, \textsc{H}e\textsc{I} {\rm and} \textsc{H}e\textsc{II}}. The frequency ranges, characteristic energies, ionisation cross-sections and dust opacities are listed in Table~\ref{tab:RTGroups}. Each of these radiation groups plays a key role in terms of radiative feedback on galaxy formation.
For each stellar particle, the mass, age and metallicity-dependent stellar specific luminosities are extracted on-the-fly from the SED model of BPASSv2.2.1 \citep{bpassv211,bpassv221} assuming a Chabrier \citep{chabrier2003galactic} IMF. 
The SED spectra (in units of $\rm erg s^{-1} \ \rm M_\odot^{-1}$ \ \AA$^{-1}$) are tabulated according to age- and metallicity-dependent luminosities for each radiation group by integrating over the energy intervals. We use this tabulated SED (single table for each simulation) to extract the number of photons emitted by each stellar particle. The photons in each group are directly injected in the host cell of the stellar particle. We update the photo-ionisation cross sections and energies at fixed intervals for the radiation groups to represent the luminosity-weighted average of all emitting sources (see \citealt{rosdahl2013ramses}). 

The photons in the UV groups not only interact with gas via photo-ionisation and photo-heating, but also via radiation pressure from photo-ionisation and dust. In contrast, the infrared and optical groups do not ionise hydrogen or helium, but interact with gas via radiation pressure on dust. 
In {\tt SPICE} we do not model dust as an active ingredient but assume that the dust number density scales with the gas-phase metallicity and the gas thermal state. Following \cite{nickerson2018simple}, we assume a local dust number density $n_{\rm d} \equiv (Z/Z_\odot) f_d n_{\rm H}$, where $Z$ is the local metallicity, $f_{\rm d} = 1 - x_{\rm HII}$ is the neutral fraction of gas that holds dust, and $n_{\rm H}$ is the hydrogen number density. This ensures that the dust density depends not only on enrichment but also on the local thermal state. While dust is generally not expected to survive in photo-ionised gas \citep{Finkelman_2012,Kannan_2020dust}, dust-to-ionised-gas ratios as high as 0.01 have been observed \citep{Harper1971,Contini_2003,Laursen2009} in ionised media. The $f_{\rm d}$ parameter in the previous equation allows for non-zero (albeit small) dust effects even in ionised media.
To each photon group, we assign dust absorption and scattering opacities as given in Table~\ref{tab:RTGroups}. If absorbed by dust, the photon flux from the group is added to the infrared group, where it will interact with gas via multi-scattered radiation pressure \citep{Costa2018a}. 
\begin{figure*}
    \centering
    \includegraphics[width=\textwidth]{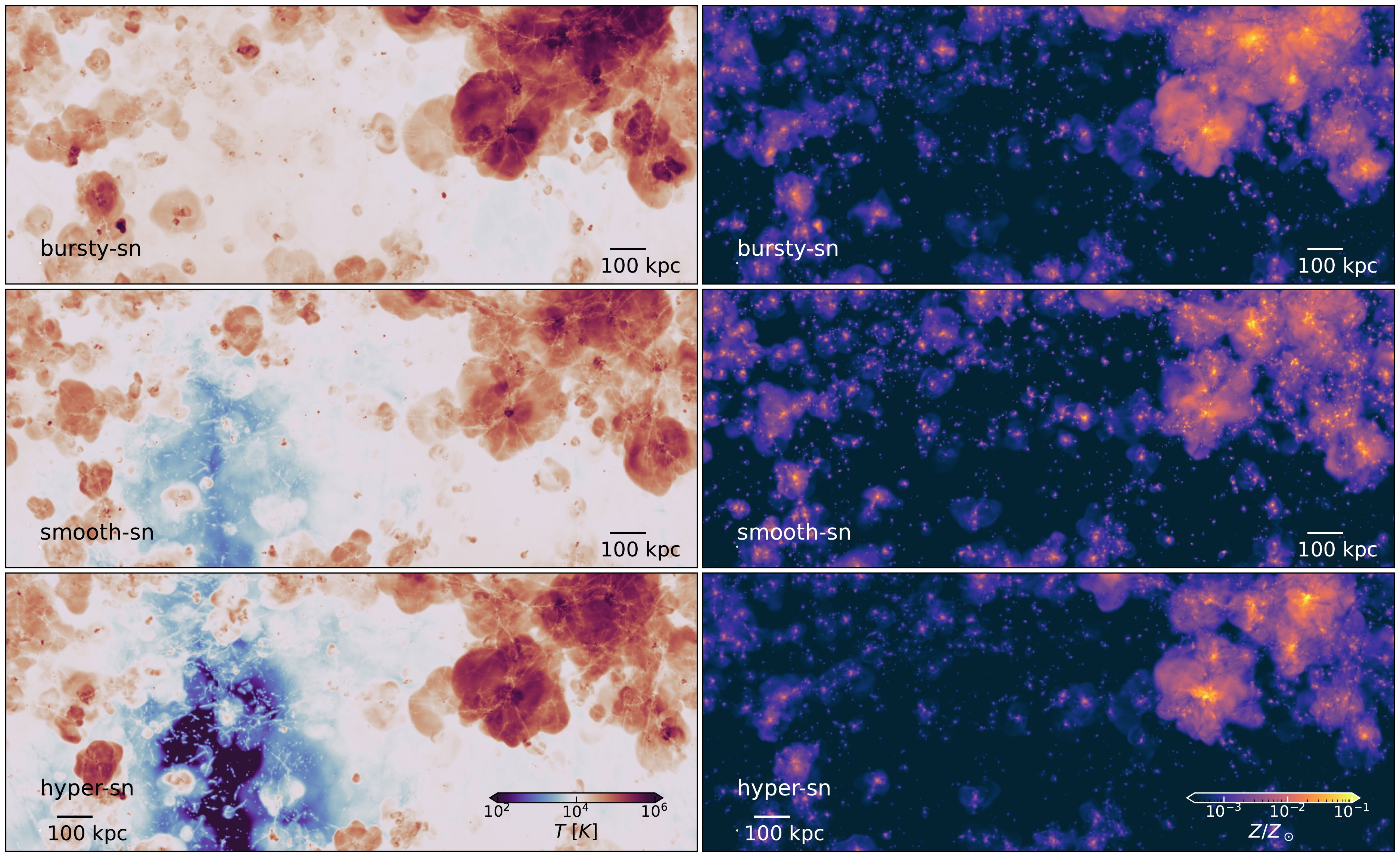}
    \caption{Temperature (left column) and gas-phase metallicity (right column) of the various simulations at $z=5$. The projections shown cover a region spanning $L_{\rm box} \times L_{\rm box}/ 3 \times L_{\rm box}$. Consequences of varying the stellar feedback are imprinted in the thermal state and enrichment of the large-scale structure.}
    \label{fig:Tmetz5}
\end{figure*}
\begin{figure}
    \centering
    \includegraphics[width=\linewidth]{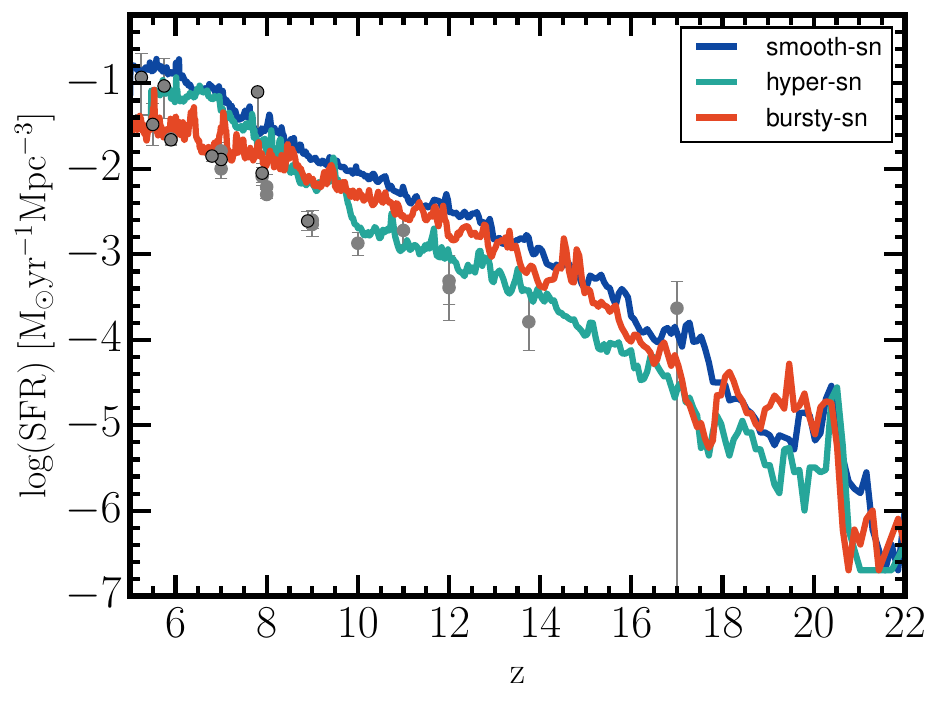}
    \caption{Cosmic star formation rate density as a function of redshift for the {\tt bursty-sn} (red line), {\tt smooth-sn} (blue) and {\tt hyper-sn} (turquoise) models. Observational constraints (see text) are marked with grey circles. \texttt{hyper-sn} most closely reproduces the observed star formation density at $z \gtrsim 10$, while \texttt{bursy-sn} is closest to observational constraints at $z \lesssim 8$.}
    \label{fig:SFRD}
\end{figure}
\section{Results}
\label{sec:results}
In this section we present our main results, starting with a qualitative overview of the simulations.
\subsection{Overview}

Primordial gas cools to form the first stellar particle at $z=22.4$. Feedback from massive stars via photo-ionisation begins immediately, and supernova feedback kicks in as early as $3 \, \rm Myr$ ($10 \ \rm Myr$) after a stellar particle is born in {\tt smooth-sn} and {\tt hyper-sn} ({\tt bursty-sn}). Galaxies continue to grow and the first $M_* > 10^7 \rm M_\odot$ dwarf galaxies form as early as $z=12$. 
In Figure~\ref{fig:miniGigaplot}, we illustrate the dynamic range of the {\tt SPICE} simulations. We note that the quantities shown in this figure are largely taken from {\tt smooth-sn}, but they can vary dramatically between models.

Panel {\it A} shows the volume-weighted gas density distribution on Mpc scales, illustrating the cosmic web at $z=5$. At this redshift, {\tt SPICE} has produced a population of $\approx 13,000, 10,000, 7,000$ resolved (see Section~\ref{sec:Halofinding}) galaxies with $M_* > 10^5 \rm M_\odot$ for the {\tt smooth-sn}, {\tt bursty-sn} and {\tt hyper-sn} models, respectively. The most massive halo at $z=5$ in each simulation has a virial mass of $2.7 \times 10^{11} \rm M_\odot$ and hosts a stellar mass of $1.9\times10^{10}\,\rm M_\odot$, $2.1\times10^9\,\rm M_\odot$ and $1.7\times10^{10} \,\rm M_\odot$ for {\tt smooth-sn}, {\tt bursty-sn} and {\tt hyper-sn}, respectively.

In panel {\tt B}, we zoom onto an over-density in the simulations and highlight the quantities that describe the effect of novel physics included in {\tt SPICE}, i.e, infrared radiation transport and consequently, radiation pressure on dust. The upper triangle in panel {\tt B} shows the acceleration imparted to gas via radiation pressure on dust, with the highest radiation pressure observed very close to the centers of dusty galaxies. The lower triangle in panel {\tt B} shows the infrared radial flux emerging from galaxies, which includes intrinsic emission from stars, dust emission and multi-scattered IR radiation. We see a network of infrared-bright galaxies with a well established background by $z=5$. 

In panel {\tt C} we zoom further into smaller scales, showing a single galaxy. We plot the radial flux of hydrogen-ionising radiation ($h \nu > 13.6 \,\rm eV$) emerging from a $M_* = 2\times10^{10} \rm M_\odot$ galaxy. We can see the escape channels of LyC radiation, with bright orange/yellow colors representing HII regions through of which ionising radiation escapes most efficiently, and blue colours represent regions of inefficient escape. We investigate the connection between galaxy morphology and escape fractions of galaxies in Section~\ref{sec:Lycfesc}. 

Panel {\tt D} shows the surface brightness of the same galaxy in {\tt [CII]} (a line which efficiently traces the cold phase of the ISM), calculated using a subgrid model which will be presented in detail in a follow-up study (Bhagwat et al, in prep). As {\tt [CII]} is sensitive to the thermal state of the ISM, which is, in turn, shaped by feedback, this line can be potentially used to constrain feedback models. 

Finally, panel {\it E} is a zoom onto a region of size $4\times R_{\rm vir}$ ($\approx 120 \rm kpc$) centred on a $\approx 2.7\times10^{11} \rm M_\odot$ halo at $z=5$. We can see this massive halo being fed by infalling gas filaments. We find further galaxies along the filaments, some with low mass companions. In {\tt E1, E2 and E3}, we show the gas distribution of the central galaxies for our three different feedback models. The {\tt bursty-sn} ({\tt E1}) produces a highly irregular and disturbed gas distribution, in contrast to the  {\tt smooth-sn} ({\tt E2}) model, which shows a well-formed disk galaxy. Finally, {\tt hyper-sn} ({\tt E3}) produces a compact disk-like density distribution. These distinct gas morphologies provide us with a first hint of the widely different effects of our feedback models on the morphology of the first galaxies.

In the left-hand panel of Figure~\ref{fig:Tmetz5}, we show the gas temperature as obtained in the three feedback models at $z=5$. The projections are calculated in a region of volume $L_{\rm box} \times L_{\rm box} / 3 \times L_{\rm box}$. 
Bright red/maroon areas trace recent hot ($T > 10^6 \, \rm K$) outflows driven by SN explosions, while the lighter brown regions ($T \approx 10^5$ K) trace older adiabatically-cooled ionised gas bubbles produced by previous star formation episodes. While these giant {\tt HII} regions are present and co-spatial in all simulations, they are hottest and largest in {\tt bursty-sn}, and least pronounced in {\tt smooth-sn}. 
Beyond the SN-driven ionised gas lies a volume-filling, diffuse component with $T \sim 10^4 \, \rm K$ gas, shown by the bright white regions. This component consists of photoionised 
material which has been irradiated by stellar radiation at distances beyond those reached by feedback-powered ionised gas. The spatial scale of this diffuse component is set by the strength of SN feedback, as SN explosions create channels through which ionising photons can escape into the low density inter-galactic medium. 
\begin{figure}
    \includegraphics[width=0.975\linewidth]{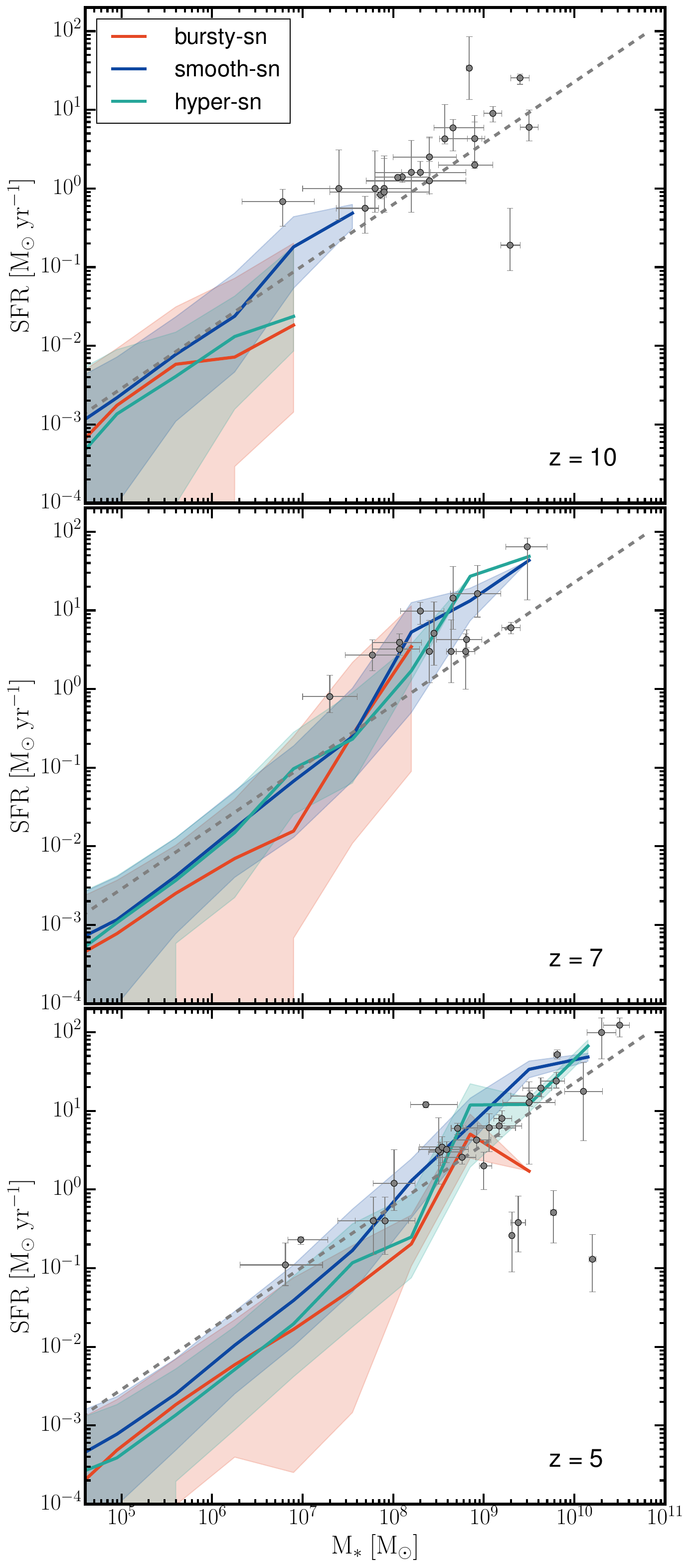}
    \caption{Star formation rate averaged over 10 Myr as a function of stellar mass for the  {\tt bursty-sn} (red line), {\tt smooth-sn} (blue) and {\tt hyper-sn} (turquoise) at $z=10$, $7$ and $5$. Solid lines represent the median SFR, with the shaded regions covering the 16th and 84th percentiles. Observations from various JWST programs (see text) are shown with grey circles. The dashed line refers to the best fit relation for the main sequence at $z=6$ \citep{Iyer_2018}.    }
    \label{fig:SFMS}
\end{figure}
A low volume-filling fraction of $T>10^4$~K gas could arise from either inefficient feedback or a smaller ionising photon budget. However, we show in Sections~\ref{sec:reionhist} and \ref{sec:Lycfesc} that the smaller volume-filling fractions seen in {\tt smooth-sn} (middle-left panel) and {\tt hyper-sn} (lower-left panel) arise in spite of a \emph{higher} photon budget and are caused by low escape fractions. We see the presence of cold voids in both {\tt smooth-sn} and {\tt hyper-sn}, with the latter being colder and more extended. The difference between these two models is solely the SN feedback, highlighting the major role this plays in reionisation. 

The right-hand panels of Figure~\ref{fig:Tmetz5} show metallicity projections for the various feedback models at $z=5$. Simulation {\tt bursty-sn} produces extended metal-enriched regions, again illustrating the stronger feedback present in this simulation. For instance, {\tt smooth-sn} shows a larger number of compact enriched gas with typically higher maximum metallicities ($Z > 1 Z_\odot$) at the very centers of halos. {\tt hyper-sn} produces a distribution similar to {\tt smooth-sn}, where the enriched regions are compact and with higher maximum metallicities. (the gas-phase metallicity of the most massive halo at $z = 5$ is $\sim0.1 Z_\odot$, $\sim0.9 Z_\odot$, $\sim0.49 Z_\odot$ for {\tt bursty-sn},{\tt smooth-sn} and {\tt hyper-sn}, respectively).
However, {\tt hyper-sn} contains visibly less structure than {\tt smooth-sn} owing to the fact that low mass galaxies are much more effectively quenched (see section~\ref{sec:starformation} and \ref{sec:luminosityfunc}). 

\subsection{Cosmic star formation history}
\label{sec:starformation}
\begin{figure*}
    \includegraphics[width=\textwidth]{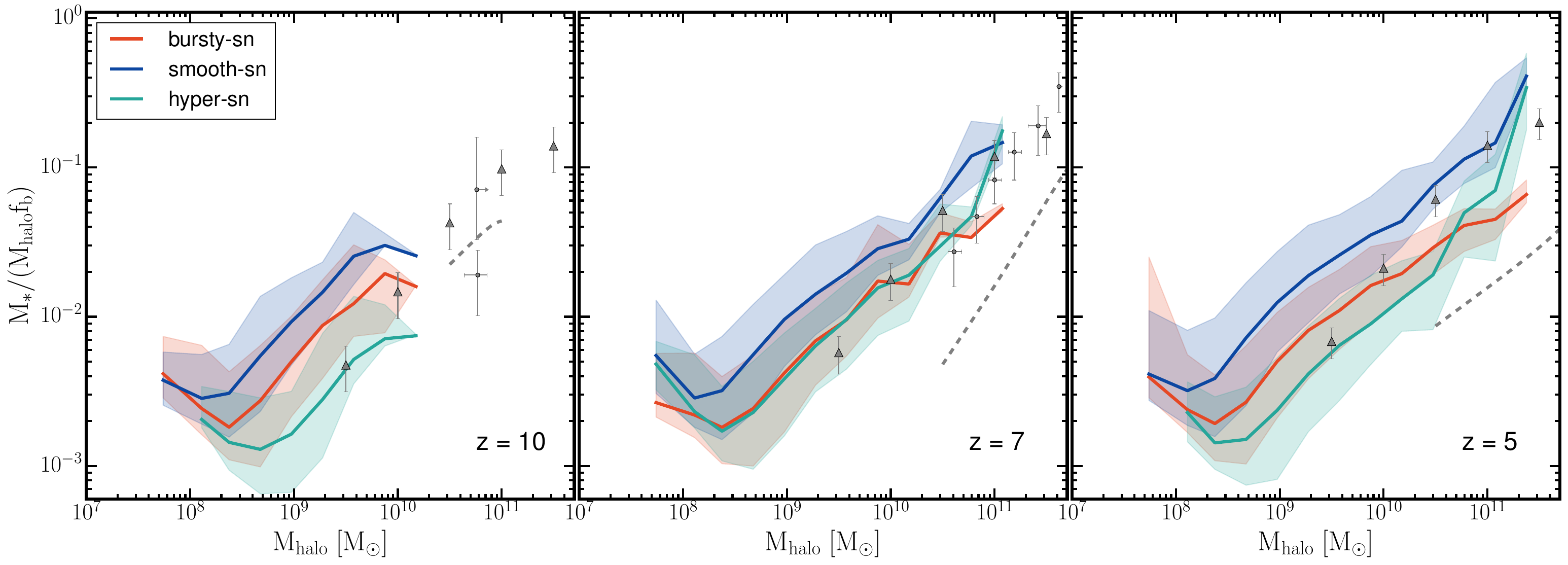}
    \caption{Star formation efficiency as a function of halo mass for {\tt bursty-sn} (red line), {\tt smooth-sn} (blue) and {\tt hyper-sn} (turquoise) at redshift $z=$ 10 (left panel), 7 (middle) and 5 (right). Solid lines represent the median SFE in each mass bin, with the shaded regions covering the 16th and 84th percentiles. 
    Grey dots are observations from \citet{Stefanon_2021}, grey triangles abundance matching estimates from \citet{Tacchella_2018abundance}, and the grey dashed line refers to the best fit from \citet{Behroozi_2019}.
    }
    \label{fig:SFE}
\end{figure*}
In Figure~\ref{fig:SFRD} we show the time evolution of the star formation rate density (SFRD) in the various feedback models calculated over the full simulation volume. We see that the SFRD increases with decreasing redshift as massive galaxies assemble. The SFRD is consistently highest in the {\tt smooth-sn} model, while it is more effectively suppressed in {\tt bursty-sn}, particularly below $z \, = \, 8$. The SFRD is initially lowest in {\tt hyper-sn} compared other models, owing to early strong feedback from HN. However, as the gas metallicity increases, the HN rate in the massive haloes is reduced exponentially (see Equation~\ref{eq:HNrate}), and the weakening feedback results in a rapid increase in star formation at $z = 9-10$. 
We also notice that the burstiness of star formation in the different models varies as mentioned in section~\ref{sec:feedbackvariations}. Both {\tt bursty-sn} and {\tt hyper-sn} have strong isolated bursts of star formation followed by relatively quiescent phases, whereas {\tt smooth-sn} shows a comparatively smooth star formation history. The strength of the bursts increases with decreasing redshift for {\tt bursty-sn}, while it decreases for {\tt hyper-sn} and {\tt smooth-sn}. 

By comparing our models to observational constraints from multi-wavelength studies \citep{2014MadauDickinson,Rowan_Robinson_2016,Khusanova_2021,Donnan_2022,YoshidaHaSFR,Harikane_2023,ALMAREBELSz7SFRD}, we see that {\tt hyper-sn} shows the best agreement at $z\gtrsim10$, {\tt bursty-sn} at $z\lesssim8$, while {\tt smooth-sn} is consistently  higher than observations.

Figure~\ref{fig:SFMS} shows the SFR of galaxies (averaged over the previous 10 Myr) as a function of stellar mass for the three feedback models and redshifts. We see that the median relations from all models roughly lie on a locus, forming a star formation main sequence (SFMS).  Although the median relation does not vary significantly across models, the scatter differs. We quantify the scatter by fitting a Gaussian to the distributions of SFRs within bins of stellar mass. At $z=10$, across the mass range, {\tt bursty-sn} shows the largest scatter ($\approx 0.8$ dex) followed by {\tt hyper-sn} ($\approx 0.5$ dex) and finally {\tt smooth-sn} ($\approx 0.3$ dex). By $z=7$, the scatter in {\tt smooth-sn} and {\tt hyper-sn} reduces to about $(0.2-0.3)$ dex, in agreement with observations from \citet{2014Speagle}. However, {\tt bursty-sn} continues to show a large scatter of $(0.8-0.9)$ dex with it being the largest below $M_* < 10^7 \rm M_\odot$. While {\tt smooth-sn} and {\tt hyper-sn} show minimal evolution in scatter between $z=7$ and $z=5$, the scatter for {\tt bursty-sn} drops to $\approx 0.4$ dex. 
 
We compare the SFMS from the three models with with observations from recent JWST programs \citep{Fujimoto_2023,haro2023spectroscopic,jung2023ceers,long2023efficient,haro2023spec,Leethochawalit_2023,Robertson_2023,heintz2023dilution,Jin_2023,helton2023jwst,atek2023jwst,Treu_2023,heintz2023extreme,Asada_2023,Bouwens_2023,looser2023jades,Papovich2023blue}. Data is collected in bins of $\pm0.25$ around the redshifts shown. At $z=10$, the vast majority of observations lie in a mass range not probed well by {\tt SPICE}, however, {\tt smooth-sn} shows the best agreement with observations. At $z=7$ and $z=5$, the median relations from all models is in excellent agreement with data for galaxies with $M_* > 10^7 \rm M_\odot$.

Figure~\ref{fig:SFE} shows the ratio between stellar mass and halo mass, which we view as a proxy for global star formation efficiency (SFE), $M_* /(M_{\rm halo} f_b)$, as a function of halo mass $ M_{\rm halo}$, where $ f_{\rm b} = \Omega_{\rm b}/\Omega_{\rm m}$. We note that all the stellar mass within the virial radius of a halo is included in the calculation. The {\tt smooth-sn} model produces the highest SFE at all times as compared to the other models, especially at the high-mass end. The {\tt bursty-sn} model has only a weakly time-dependent SFE, while the SFE in {\tt hyper-sn} starts at very low values at early times, but, following a starburst episode at $z \approx 9$, increases, and by $z=7$ becomes comparable to the one of the {\tt bursty-sn} model. The HN rate in haloes with $M_{\rm halo} >10^{10} \,\rm M_\odot$ reaches 1\% (by $z\approx9$), therefore, the energetic component of feedback becomes negligible\footnote{The range $z=9-10$ also marks the onset of galaxies settling into disks. We plan to quantify this morphological evolution in a follow up work.}. Consequently, the SFE in these haloes increases rapidly due to inefficient feedback. Lower-mass halos have small SFEs, likely because the high HN rate combined with the shallower potential wells lead to more effective quenching. All three models show consistently higher ($\approx0.6-1$ dex) SFEs as compared to best fit estimate from \citet{Behroozi_2019}. They are however in agreement with abundance matching estimates from \citet{Tacchella_2018abundance}, and $z$=7 observations from \citep{Stefanon_2021}. We note that the trend reversal at low halo masses is due to these halos undergoing their first starburst events (average stellar age $\sim 5-15$ Myr), which temporarily boosts their SFEs.

Overall, the results shown in this section highlight how some observables (e.g SFRD, SFE) vary strongly with implementation of feedback, while others (e.g SFMS) are less affected. The explosiveness of the {\tt bursty-sn} model highlights the need for sustained strong feedback to maintain a sufficiently low star formation rate. In contrast, the {\tt smooth-sn} model, characterized by less energetic feedback, struggles to effectively regulate star formation, thus allowing for the formation of massive star-forming galaxies by $z=5$. Finally, the {\tt hyper-sn} model emphasizes the importance of striking a balance between bursty and smooth feedback mechanisms. 

\subsection{Reionisation histories}
\label{sec:reionhist}
\begin{figure*}
    \includegraphics[width=\textwidth]{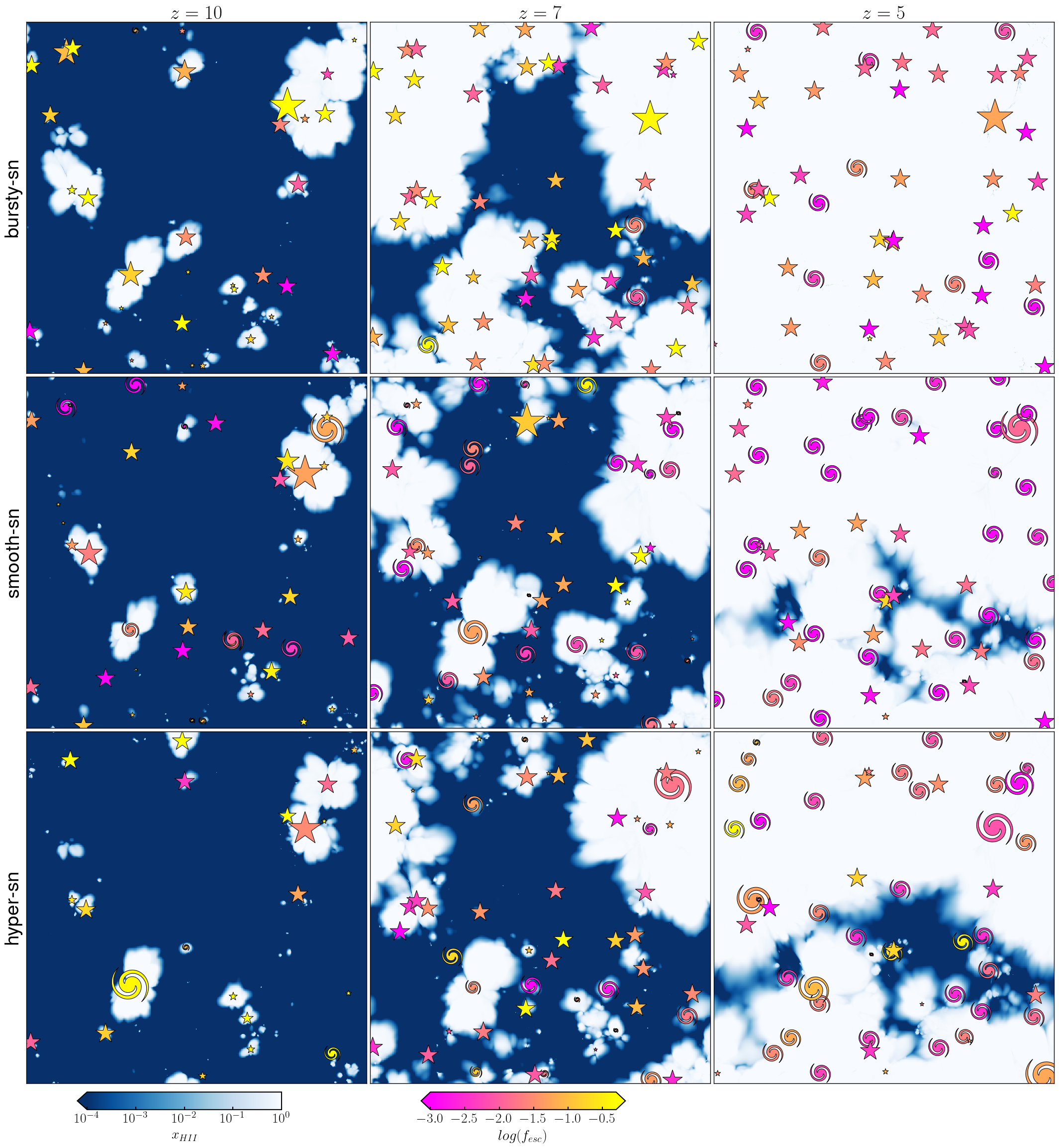}
    \caption{Thin-slice projections showing HII fraction in the {\tt bursty-sn}, {\tt smooth-sn} and {\tt hyper-sn} models (from top to bottom) at $z=10$, $7$ and $5$ (from left to right). Locations of galaxies are over-plotted (see text for selection criteria), with spiral (star) symbols representing rotationally- (dispersion-) dominated galaxies (gas morphologies; see section~\ref{sec:morphologies}). Each symbol is color coded with the LyC escape fraction of the respective galaxy, with the sizes of markers representing escaping ionising photon luminosity. 
    Already at $z \, = \, 10$ ionised regions show differences in sizes, with {\tt bursty-sn} and {\tt hyper-sn} producing the largest and smallest, respectively. By $z \, = \, 5$, reionisation is complete in {\tt bursty-sn}, whereas significant neutral patches persist in the other two models.}
    \label{fig:reion-proj}
\end{figure*}

\begin{figure}
    \includegraphics[width=\linewidth]{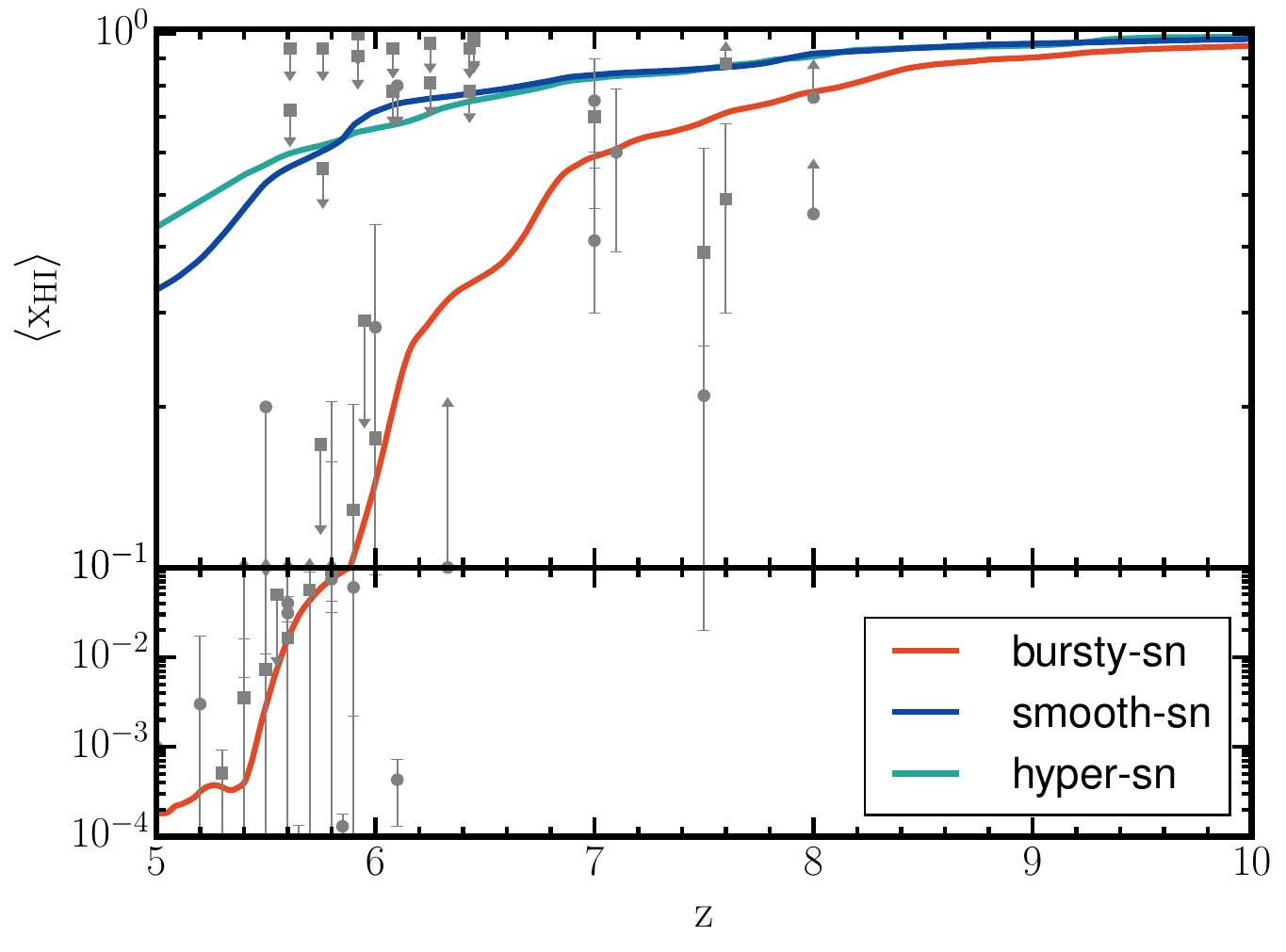}
    \caption{Volume-weighted neutral hydrogen fraction as a function of redshift for the three feedback models. Observational data shown in grey points are taken from \citet{Fan2006,McGreer2011,2012Ono,Schroeder_2012,McGreer2015,Grieg2017,2018Mason,2019Mason,Greig_2019,Hoag_2019,Yang_2020,2020Wang,2020Lu,2020Jung,2021Choudhury,2022Bosman,2022Zhu,2023Gaikwad}.
    {\tt bursty-sn} reionises within observational constraints whereas {\tt smooth-sn} and {\tt hyper-sn} reionise late.}
    \label{fig:ReionHist}
\end{figure}
\label{sec:reionisation_history}
In Figure~\ref{fig:reion-proj} we show thin-slices (depth of 100 ckpc) of ionised fractions ($x_{\rm HII}$) in the three feedback models representing the progress of reionsation, i.e, formation of ionised regions at high redshifts, overlap and post-overlap phase. The difference in their sizes and abundance is already evident at $z=10$, when {\tt bursty-sn} produces generally larger HII regions, while the {\tt smooth-sn} and {\tt hyper-sn} models show a combination of extended and localised regions, with {\tt hyper-sn} exhibiting a larger number of small ones. The differences are amplified at $z=7$, where we observe that the bubbles in {\tt bursty-sn} are already deep into the overlap phase, with islands of neutral hydrogen localised in voids with small volume-filling fractions, while both {\tt smooth-sn} and {\tt hyper-sn} still show mostly isolated HII regions with large neutral voids in between. 
By $z=5$, the overlap phase is completed in {\tt bursty-sn}, with neutral hydrogen present only in dense self-shielded structures, while in both {\tt smooth-sn} and {\tt hyper-sn} small neutral islands still persist. 

To demonstrate the varied ability of galaxies to reionise their surroundings, in Figure~\ref{fig:reion-proj} we divide each projection in a $8\times8$ grid and mark the location of the most luminous galaxies within each grid sub-division. The galaxies marked within each ionised bubble account for $>95\%$ of the escaping ionising photon budget, effectively representing the galaxies that are responsible for the formation and growth of such bubble.
Each galaxy is coded by its morphology (stars and spirals represent dispersion- and rotation-dominated galaxies, respectively; see section~\ref{sec:morphologies} for details), LyC escape fraction (different colors) and escaping ionising luminosity (marker sizes). In Section~\ref{sec:Lycfesc}, we discuss in more detail the connection between galaxy morphology and LyC escape fractions.

In Figure~\ref{fig:ReionHist} we show the resulting reionisation histories. We find that while {\tt bursty-sn} reionises\footnote{We consider reionisation to be complete if the volume-averaged neutral hydrogen fraction of the simulation reaches $10^{-4}$.} the Universe by $z \, = \, 5.1$, this is not the case for {\tt smooth-sn} and {\tt hyper-sn} models. In {\tt bursty-sn} the reionisation process starts early, with the hydrogen neutral fraction, $\rm x_{HI}$, dropping to $\approx 0.9$ already at $z \approx 10$. From $z \, \approx \, 7$, $\rm x_{HI}$ experiences a steep decline in a series of steps that are time-correlated with three star formation bursts clearly visible in Figure~\ref{fig:SFRD}, and resulting in a `late reionisation scenario' \citep{Kulkarni19,Keating20,2023Gaikwad}, with $x_{\rm HI}$ becoming lower than $10^{-4}$ at $z \approx 5.1$. Despite producing more star-forming galaxies (see Section~\ref{sec:starformation}), {\tt smooth-sn} and {\tt hyper-sn} result in a much later reionisation, failing to bring the volume-weighted neutral fraction below $ 30 \%$ even by $z \, = \, 5$, the time when our simulations end. 
However, we note that the simulation volumes are not large enough to derive a converged mean reionisation history \citep{Iliev_2014,gnedin2022modeling}, and a box size of $\gtrsim 100$~cMpc/$h$ is required to better represent an average region of the universe and to account for the typical sizes of ionised regions in the late stages of reionisation, as these can be tens of cMpc in size. 

The stellar masses integrated over the entire simulation volume at $z=5$ for the {\tt smooth-sn}, {\tt bursty-sn} and {\tt hyper-sn} model are $\approx 5 \times 10^{10}$~M$_\odot$, $\approx 2 \times 10^{11}$~M$_\odot$ and $\approx 1 \times 10^{11}$~M$_\odot$, respectively (a ratio of $1:4:2$). These estimates imply a different number of SN explosions and intrinsic ionising photon budget, as both increase for larger stellar masses. Therefore, {\tt smooth-sn} and {\tt hyper-sn} have a larger ionising photon budget as compared to {\tt bursty-sn}, suggesting that the late reionisation is not a result of a lower ionising photon budget, but rather it is due to inefficient escape of photons. 

Explosiveness of feedback affects star formation which determines the ionising photon budget and the extent to which gas is disturbed (modulation of escape fractions). {\tt smooth-sn} and {\tt hyper-sn} have a weaker feedback (especially at $z<8$), which allows for the formation of stable gas configurations (see Figure~\ref{fig:miniGigaplot}) and the overproduction of stars (see Figures~\ref{fig:SFRD},\ref{fig:SFE} and Section~\ref{sec:starformation}), but it is unable to significantly disrupt the gas configuration, hence suppressing the escape of radiation. 

\subsection{Luminosity functions at 1500 \AA}
\label{sec:luminosityfunc} 

\begin{figure}
    \includegraphics[width=\linewidth]{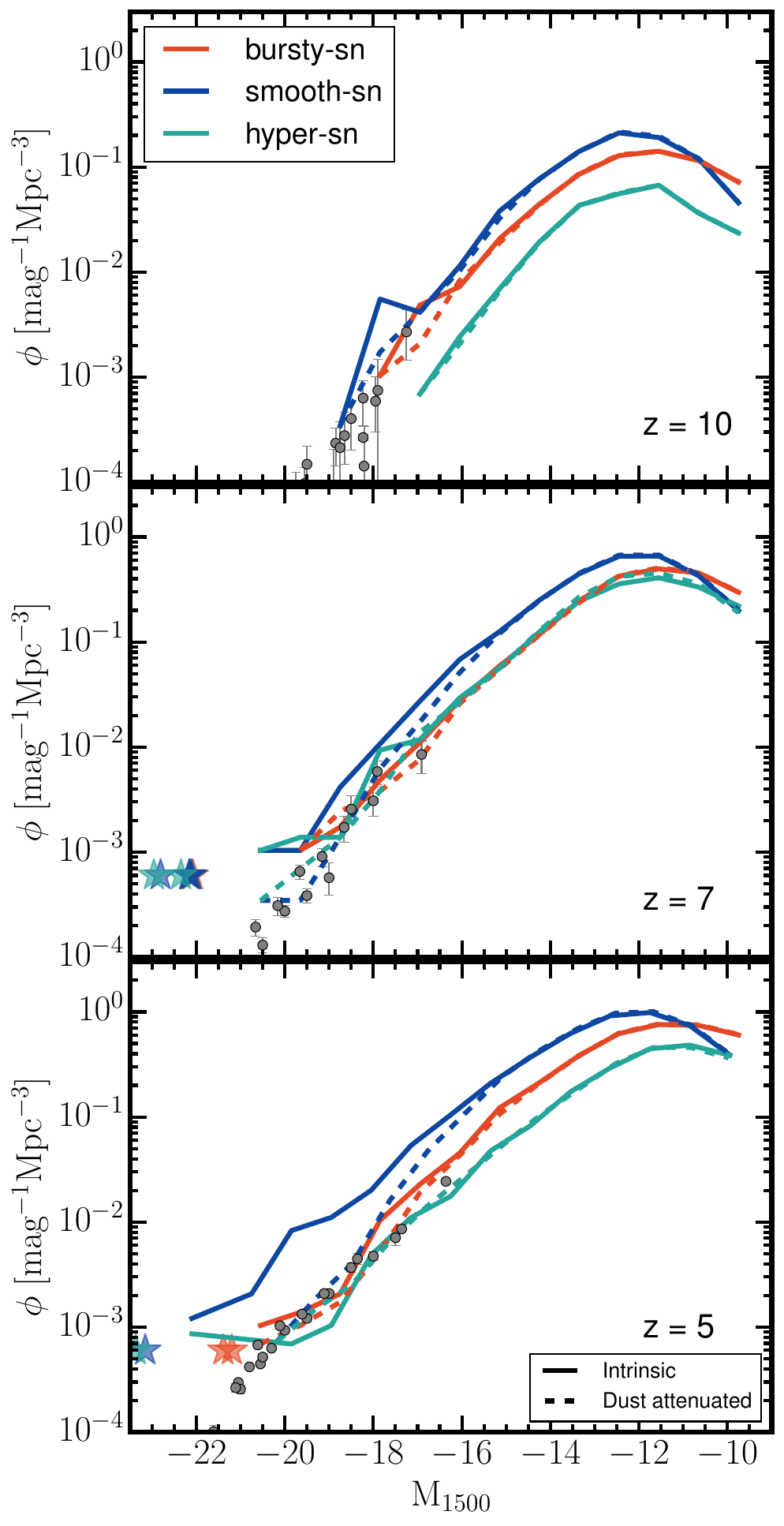}
    \caption{1500~\AA~luminosity functions (LF) for the various feedback models at $z=10$, $7$ and $5$ (from top to bottom). The solid and dashes lines refer to the intrinsic and  dust-attenuated luminosity functions, respectively. Observations from HST and JWST fields are marked in grey (see text). Stars represent intrinsic luminosities in bins with fewer than $3$ galaxies. Despite intrinsic LFs being different, dust-attenuated ones show minimal differences between models. Therefore, LFs \emph{cannot} be used to constrain feedback models.}
    \label{fig:UVLF}
\end{figure}

In Figure~\ref{fig:UVLF} we show the evolution of the 1500 \AA~luminosity function, where the luminosities are calculated in a 10 \AA \ bin around 1500 \AA \ using the stellar SEDs. 
Solid and dashed lines refer to the intrinsic and dust-attenuated luminosity functions, respectively. The latter is calculated with the Monte Carlo line transfer code  {\tt RASCAS} \citep{Michel_Dansac_2020}, by casting 100 rays from each stellar particle within a galaxy to the edge of the halo, and evaluating the solid angle-averaged attenuation per galaxy. The orientations of the rays are sampled randomly, and the dust attenuation along each ray is calculated using the dust model described in Section~\ref{sec:StellarFeedback}. We also show observational estimates of the luminosity functions from HST legacy fields and recent JWST programs \citep{Finkelstein_2015,Bouwens_2015,Harikane_2022,mcleod2023galaxy,adams2023epochs,Harikane_2023,Bouwens_2023,Bouwens_2023uv,leung2023ngdeep}.    

\begin{figure*}
    \centering    
        \includegraphics[width=\textwidth]{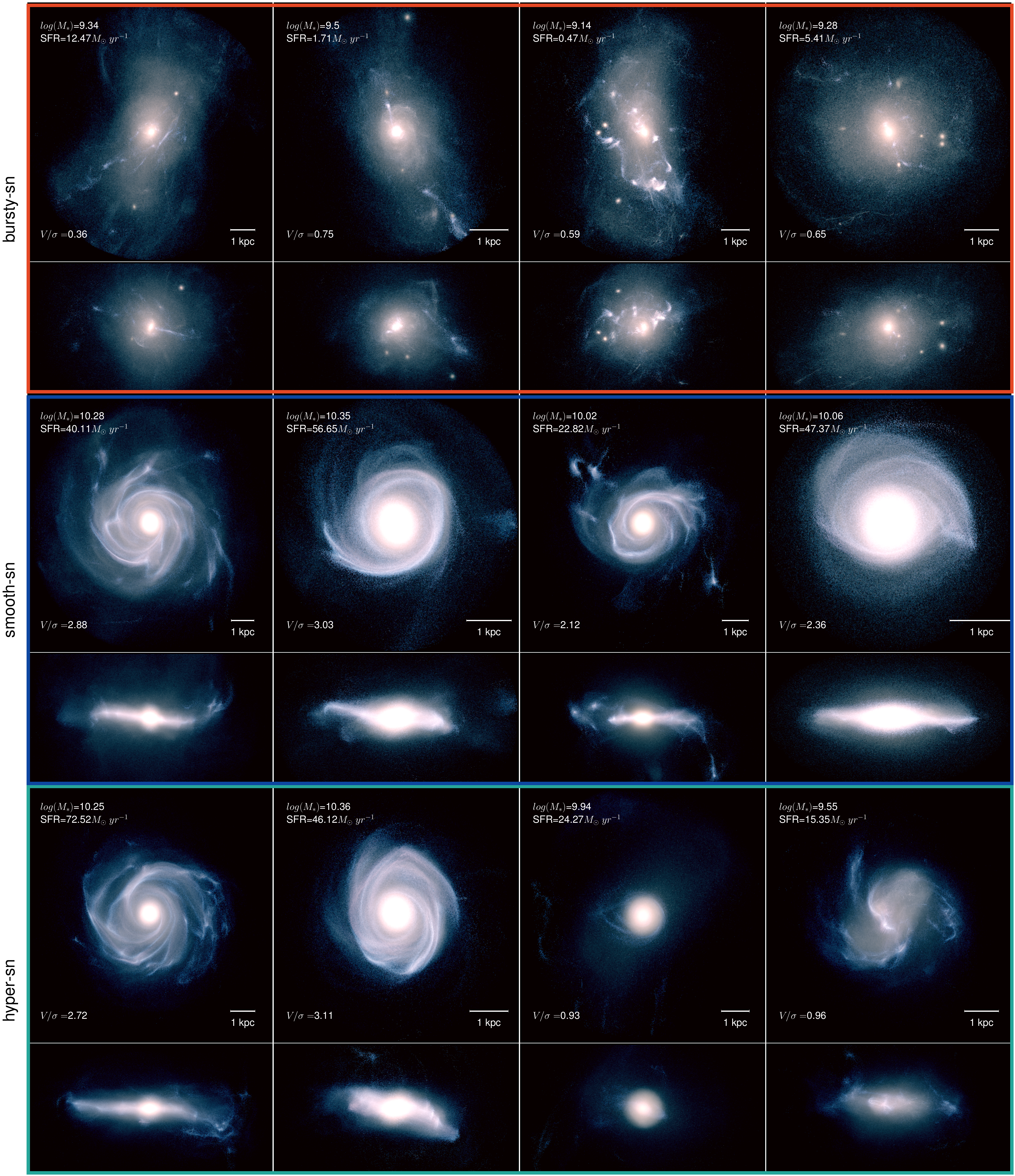}
    \caption{The columns show JWST RGB composite projections (face-on and edge-on) in the F200W, F277W and F444W filters for the four most massive galaxies at $z \, = \, 5$ in the {\tt bursty-sn} (top two rows), {\tt smooth-sn} (middle two rows), and {\tt hyper-sn} (bottom two rows) model. Numbers in each panel indicate the stellar mass, SFRs and $V/\sigma$ values for the stellar component. The {\tt smooth-sn} model produces mainly rotationally supported, massive disk galaxies that are bulge-heavy; the {\tt bursty-sn} model dispersion-supported, disturbed systems which are redder and less-massive; the {\tt hyper-sn} model a combination of massive rotatinally-supported spiral galaxies and dispersion-supported ellipsoids. Stellar feedback processes profoundly affect galaxy kinematics, colours and morphology.
    }
    \label{fig:Galaxyshape}
\end{figure*}

From the top panel of Figure~\ref{fig:UVLF} we see that at $z=10$ {\tt hyper-sn} produces fewer galaxies at all magnitudes compared to the other two models. The latter have similar luminosity functions, with {\tt smooth-sn} producing the highest number of (intrinsically) bright objects. Following a decline of the HN rate in massive haloes in {\tt hyper-sn}, the increase in star formation boosts the intrinsic LF, which becomes almost indistinguishable from the LF in the {\tt bursty-sn} model at $z=7$. Similarly to $z=10$, {\tt smooth-sn} shows a larger number (by 0.3-0.4 dex) of galaxies at all magnitudes compared to the other models. 

By $z=5$, {\tt smooth-sn} produces extremely UV-bright ($M_{1500} \sim -23$) objects and an abundance of galaxies at all magnitudes. Also {\tt hyper-sn} results in a similar abundance of objects at $M_{1500} \sim -23$, while showing less dimmer galaxies ($-20 < M_{1500} < -12)$. Meanwhile, {\tt bursty-sn} lies in between the other two models. 

At $z=10$ the dust-attenuated LFs are similar to the intrinsic ones for {\tt bursty-sn} and {\tt hyper-sn}, while we observe significant attenuation for  {\tt smooth-sn} at $M_{1500} < -17$. Dust attenuation becomes more significant at $z=7$ for {\tt smooth-sn} and {\tt hyper-sn}, when an effect is visible at the bright end of the LF. Indeed, it becomes relevant at $M_{1500} < -16$ for {\tt smooth-sn}, and at $M_{1500} < -18$ for {\tt hyper-sn}, while in {\tt bursty-sn} dust attenuation is minor even at the brightest end of the LF. At $z=5$, dust attenuation for {\tt smooth-sn} is significant ($>1$ dex) at $M_{1500} < -18$, while in {\tt hyper-sn} it is present only at $M_{1500} < -20$. Finally, for {\tt bursty-sn} it is minimal ($\approx 0.2 {\rm dex}$) for $M_{1500} < -17$. 
At $z=10$, while {\tt bursty-sn} and {\tt smooth-sn} agree very well with data, {\tt hyper-sn} shows a deficit of bright galaxies. However, both at $z=7$ and $z=5$, dust-attenuated LFs from all three models are in excellent agreement with observations for $M_{1500} < -16$. 

Overall, in this section we show that despite the intrinsic LFs being extremely different, the dust-attenuated ones are very similar below $M_{1500} < -16$, suggesting that UVLFs cannot be directly used to probe stellar feedback at high redshifts. 
\begin{figure*}
    \centering    
    \includegraphics[width=\textwidth]{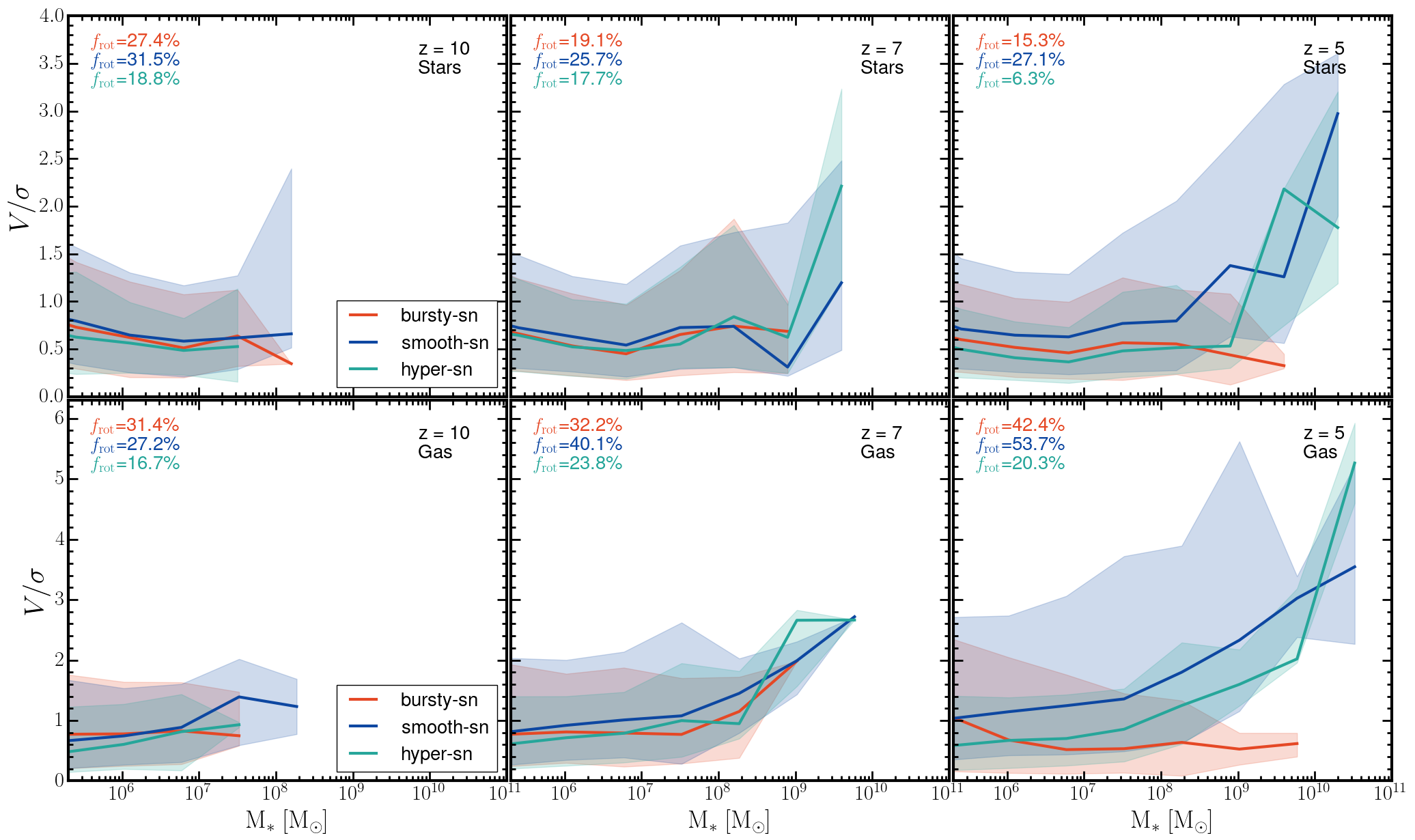}
     \caption{Median $V/\sigma$ values for stars (top row) and gas (bottom row) as a function of stellar mass for galaxies in the various feedback models at $z=10,7$ and $5$ (from left to right). 
     The shaded regions show the 16th and 84th percentiles. Percentages in each panel refer to the number of rotationally-supported galaxies at a given redshift. Large differences emerge in the kinematics by $z=5$, with {\tt smooth-sn}, {\tt bursty-sn} and {\tt hyper-sn} producing mainly galaxies which are rotationally-supported, dispersion-supported, and mixed, respectively. 
    }
    \label{fig:VSigma}
\end{figure*}
\subsection{Galaxy morphologies}
\label{sec:morphologies}
Figure~\ref{fig:Galaxyshape} shows stellar light projections for the four most massive galaxies produced in each feedback model, which are matched across simulations using the IDs of the dark matter particles comprising their parent haloes (which remain identical across simulations). The images consist of RGB composites in the F200W(B) + F277W(G) + F444W(R) JWST filters. Each galaxy is shown both face-on (upper rows) and edge-on (bottom rows), where the total angular momentum vector was used to define the orientation. Note that dust attenuation is not taken into account, i.e. the images illustrate intrinsic stellar emission only. 

The most massive galaxies in {\tt smooth-sn} are blue, spiral galaxies (see Figure~\ref{fig:Galaxyshape}), which host also bright, red bulges, and have  star formation rates reaching $\approx 50 \, \rm \rm M_\odot \, yr^{-1}$. Within the same haloes, {\tt bursty-sn} generates systems with 10 times lower star formation rates and $(0.8 \-- 1) \, \rm dex$ lower stellar mass. The host galaxies now show no evidence of disks, and appear much more irregular, hosting a number of stellar clumps and streams. Finally, the {\tt hyper-sn} model results in a mixture of blue star-forming spirals with SFRs of $\approx 50 \, \rm \rm M_\odot \, yr^{-1}$, similar to those of {\tt smooth-sn}, and irregular galaxies which also tend to be star-forming, unlike those in {\tt bursty-sn}. Variations in supernova feedback have a particularly striking impact on galaxy morphology, with differences that are systematic and persist down to stellar masses of $\approx 10^7 \, \rm \rm M_\odot$ (see Figure~\ref{fig:VSigma}).

While in Figure~\ref{fig:Galaxyshape} we show the strong qualitative differences caused by feedback on the four most massive galaxies, we also quantify them for the entire galaxy population using the ratio $V/\sigma$ \citep{Dubois_2016,Pillepich2019TNG50} as a proxy for morphology,  where $V$ is the rotational velocity and $\sigma$ is the 3D velocity dispersion. This ratio classifies the degree to which a galaxy is supported by dispersion or rotation. 
We calculate $V/\sigma$ for both stars and gas for each galaxy with $M_*>10^5$~M$_\odot$. 
For stars, we construct a velocity dispersion radial profile, as well as the 3D rotation curve for all stars within twice the stellar half mass radius of each galaxy\footnote{We bin the rotation and dispersion curves at a fixed bin-width of 0.2 ckpc ($\sim$8 times the maximum resolution of the simulation) to ensure statistical mean/medians in each bin.}. As for gas using all cells inside a fixed radius can lead to noisy calculations due to inflows and outflows,  we select gas cells within $R_{\rm vir}$ adopting an H$\alpha$ emissivity threshold. Indeed, recombination lines of atomic hydrogen from the ionised ISM are often used to measure gas kinematics in a wide redshift range \citep{degraaff2023ionised}.  We evaluate the case-B recombination volume emissivity for each gas cell as \citep{Hummer1987}: 
\begin{equation}
    \epsilon_{{\rm H}\alpha} = h\nu \ P_{\rm B}(T) \ \alpha_B(T) n_{\rm e} n_{\rm p},
\end{equation}
where $\alpha_{\rm B} = 2.753\times10^{-14} \ \lambda^{1.5}/(1.0 + (\lambda/2.74)^{0.407})^{2.242}$ $\rm cm^3 s^{-1}$ with $\lambda = 315614/T$ is the case-B recombination coefficient \citep{Hui1997}, $T$ is the gas temperature, $n_{\rm e}$ and $n_{\rm p}$ are the number densities of electrons and protons, and $P_{\rm B}(T)$ is the conversion probability per recombination event, which is $\approx 0.451$ for $T=10^4$~K. 
 
To calculate gas kinematics we select gas cells with $\epsilon_{{\rm H}\alpha}>\rm 10^{-6} \ erg \ s^{-1}\ cm^{-3}$, noting that this choice does not affect our qualitative results\footnote{This limit is taken using results on completeness estimates and theoretical studies from \citet{Belfiore_2022} and \citet{Tacchella_2022}}.
Following the process used in various theoretical and observational studies \citep{Pillepich2019TNG50,2020Rizzo}, $V/\sigma$ is calculated using $V$ at the peak of the rotation curve and $\sigma$ as the mean velocity dispersion. We classify galaxies with $V/\sigma \geq 1$ (for both stars and gas) as rotationally-supported, otherwise as dispersion-supported. 

In Figure~\ref{fig:VSigma} we show the median $V/\sigma$ for stars (top row) and gas (bottom) as a function of stellar mass. The numbers in the various panels give the fraction of all galaxies that are rotationally-supported systems at a given redshift. The stellar component in {\tt bursty-sn} and {\tt hyper-sn} shows a consistent decrease with decreasing redshift in the fraction of rotationally-supported galaxies ({\tt hyper-sn} exhibits a sudden drop from $\approx 18\%$  at $z=7$ to $\approx 6\%$ $z=5$), whereas the fraction does not change significantly in {\tt smooth-sn}. The gas component behaves differently for all models. {\tt bursty-sn} shows a steady increase in the fraction of rotationally-supported galaxies with decreasing redshift (from $\approx 31\%$ at $z=10$ to $\approx 42\%$ at $z=5$), {\tt smooth-sn} has a similar trend but the increase is more drastic (from $\approx 27\%$ to $\approx 53\%$), while {\tt hyper-sn} shows a mild increase in this fraction between $z=10$ and $7$, followed by a drop in the range $z=7-5$.  

At $z=10$, both stellar and gas components are indistinguishable between models. Similarly at $z=7$, with the only differences appearing at the highest masses, where {\tt smooth-sn} and {\tt hyper-sn} show formation of highly rotationally-supported systems. The differences between the various models become most pronounced at $z=5$ for both stellar and gas components. {\tt smooth-sn} shows a population of highly rotationally-supported galaxies ($V/\sigma >2$) and very large scatter at $M_* \gtrsim 10^8 \rm M_\odot$, while {\tt bursty-sn} produces a dispersion-supported galaxy population with a small scatter around the median. The {\tt hyper-sn} model exhibits a transition at $M_* \approx 10^8 \rm M_\odot$: scatter is small and galaxies are mainly dispersion-supported for masses lower than this value, while more massive galaxies are typically rotationally-supported and scatter is more significant. 
\begin{figure}
    \includegraphics[width=\linewidth]{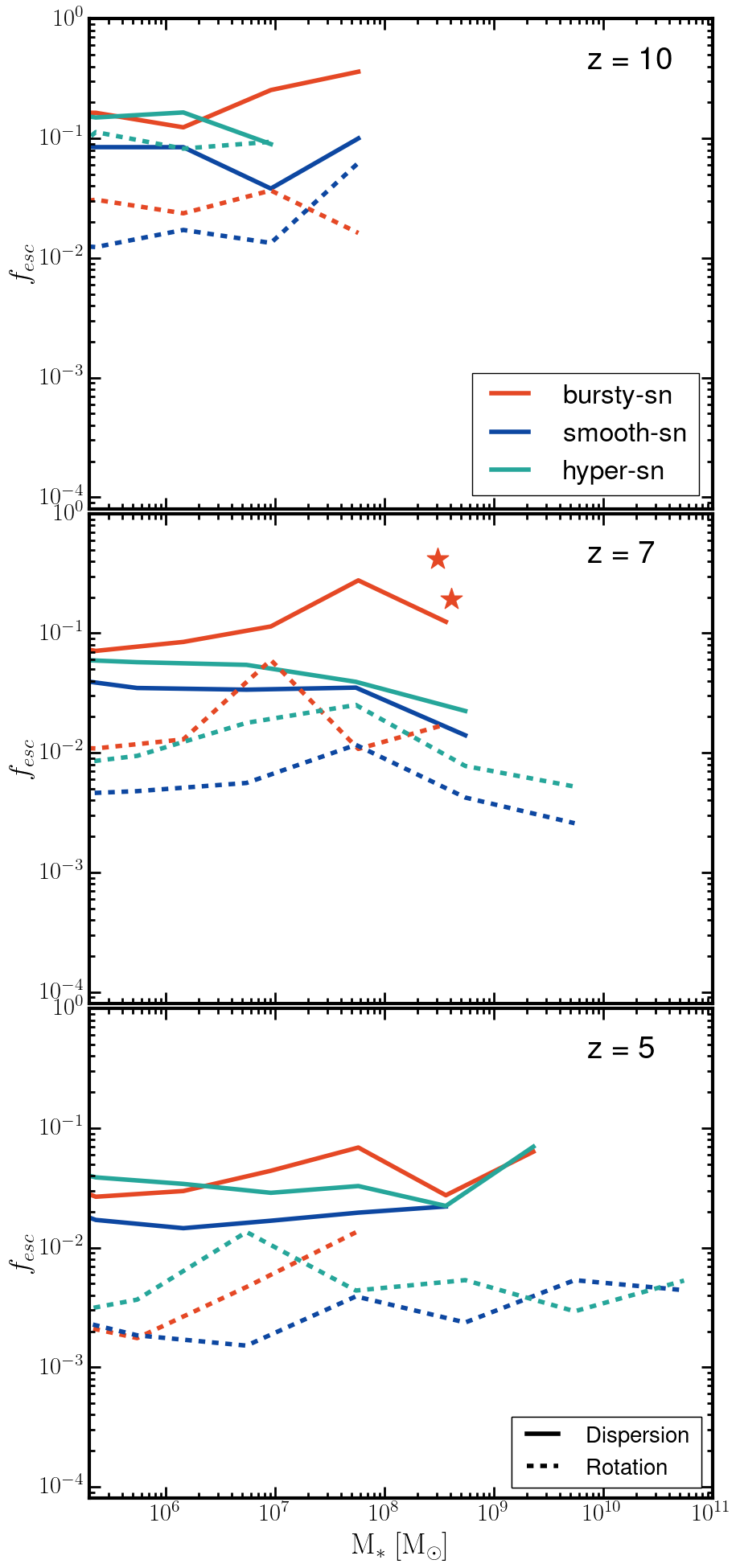}
    \caption{Luminosity-weighted mean escape fractions of {\tt SPICE} galaxies as a function of stellar mass. Solid and dashed lines refers to dispersion- and rotation-dominated systems \textbf{($V_{\rm gas}/\sigma_{\rm gas}$)}, respectively. Solid (hollow) stars represent single dispersion-(rotation-) dominated galaxies in bins with fewer than $3$ galaxies.}
    \label{fig:fescvsMstar}
\end{figure}
We note that the gaseous component shows a larger degree of rotational support as compared to the stellar component (i.e. $V_{
\rm star}/\sigma_{\rm star} < V_{\rm gas}/\sigma_{\rm gas}$), a trend particularly prominent in the {\tt smooth-sn} and {\tt hyper-sn} models.  The number of galaxies classified as rotationally- vs dispersion-supported is also widely different for the three models, especially at $z=5$, implying the presence of kinematic misalignment between the stellar and gas components in the galaxies. 

Examining the galaxy morphologies, we show that feedback plays a key role in shaping the morphological mix of galaxies that emerge post-reionisation. The populations are hard to distinguish using $V/\sigma$ as an indicator at $z>7$, except for the most massive galaxies. Below this redshift however, the morphological mixes diverge and hence can act as an indicator to distinguish feedback models. 

\subsection{Implications for LyC escape : What galaxies drive reionisation?}
\label{sec:Lycfesc}

LyC escape can proceed either through low density channels created by stellar feedback, or if the ISM is highly ionised and optically thin  \citep{Zackrisson_2013,Katz_2022}. These scenarios are not mutually exclusive and previous studies \citep{trebitsch2017fluctuating,kimm2017feedback,SphinxRosdahl2018,Barrow2020,Katz_2022,2023Yeh} show that LyC escape fraction 
($f_{\rm esc}$) is strongly regulated by feedback through a complex multiphase ISM. Therefore, the morphology of the ISM holds signatures that can aid in the identification of the galaxies responsible for reionisation.

Here we connect the morphologies of galaxies as discussed in the previous section to their $f_{\rm esc}$, which is calculated using {\tt RASCAS} \citep{Michel_Dansac_2020}. Photon packets are injected at the position of each stellar particle with a probability proportional to its LyC luminosity, evaluated using the stellar SEDs. We set the number of photon packets per halo to be 100 times the number of star particles, with a maximum of $10^7$. Photons are propagated until they are absorbed or reach a distance equal to the virial radius of the host halo. The escape fraction is then defined as the ratio between the number of escaping and injected photons.

We assign a LyC $f_{\rm esc}$ to each galaxy, and use $V_{\rm gas}/\sigma_{\rm gas}$ (as calculated in section~\ref{sec:morphologies}) to sort galaxies into rotationally- and dispersion-supported categories. In Figure~\ref{fig:fescvsMstar} we show the luminosity-weighted mean LyC escape fractions as a function of stellar mass for dispersion- (solid lines) and rotationally-supported  (dashed) galaxies. 
We observe a clear difference in the $f_{\rm esc}$ of rotation- and dispersion-dominated systems, with the latter exhibiting $\approx10-50\times$ higher escape fractions as compared to their rotational counterparts. This trend holds for all three feedback models at all redshifts in consideration. The differences in $f_{\rm esc}$ imply that at the same intrinsic luminosity and in the same environment, dispersion-dominated systems produce ionised bubbles larger than those of their rotation-dominated counterparts. 
The relative difference between the $f_{\rm esc}$ for rotation- vs dispersion- dominated systems is the largest for {\tt smooth-sn}, followed by {\tt bursty-sn} and {\tt hyper-sn}. However, the morphological mix in the three models is redshift and stellar mass dependent. Since dispersion-dominated systems are more prevalent if stellar feedback is more explosive, reionisation occurs earlier in {\tt bursty-sn} as compared to the other two models, where rotation-dominated systems are preferentially produced, especially at the highest masses. 
In Figure~\ref{fig:reion-proj} we show the connection between galaxy morphologies, escape fraction and their resulting ionised bubbles. At $z=10$ {\tt smooth-sn} and {\tt hyper-sn} exhibit a mix of star symbols, i.e dispersion-dominated systems, and spiral symbols, i.e rotation-dominated systems, while the latter dominate by $z=5$. Differently, {\tt bursty-sn} is dominated by star symbols (i.e. dispersion-dominated systems) at all redshifts. The morphological trends are also reflected in the $f_{\rm esc}$, where {\tt smooth-sn} and {\tt hyper-sn} largely show small escape fractions (pink/orange symbols), whereas in {\tt bursty-sn} larger values are more prominent  (yellow/orange symbols). We note that particularly at $z=7$, the largest ionised bubble contains a network of yellow star symbols (i.e. strong LyC leakers which are dispersion-dominated) as in {\tt bursty-sn}, whereas bubbles in which there are more pink spiral symbols (i.e. rotation-dominated systems with low escape fractions) are smaller in size. By $z=5$, the strong LyC leakers (star symbols) in {\tt bursty-sn} are able to completely reionise their surroundings, whereas this is not the case for the rotation-dominated galaxies in {\tt smooth-sn} and {\tt hyper-sn}  (which tend to be massive, see Figure~\ref{fig:VSigma})
\begin{figure}
    \includegraphics[width=\linewidth]{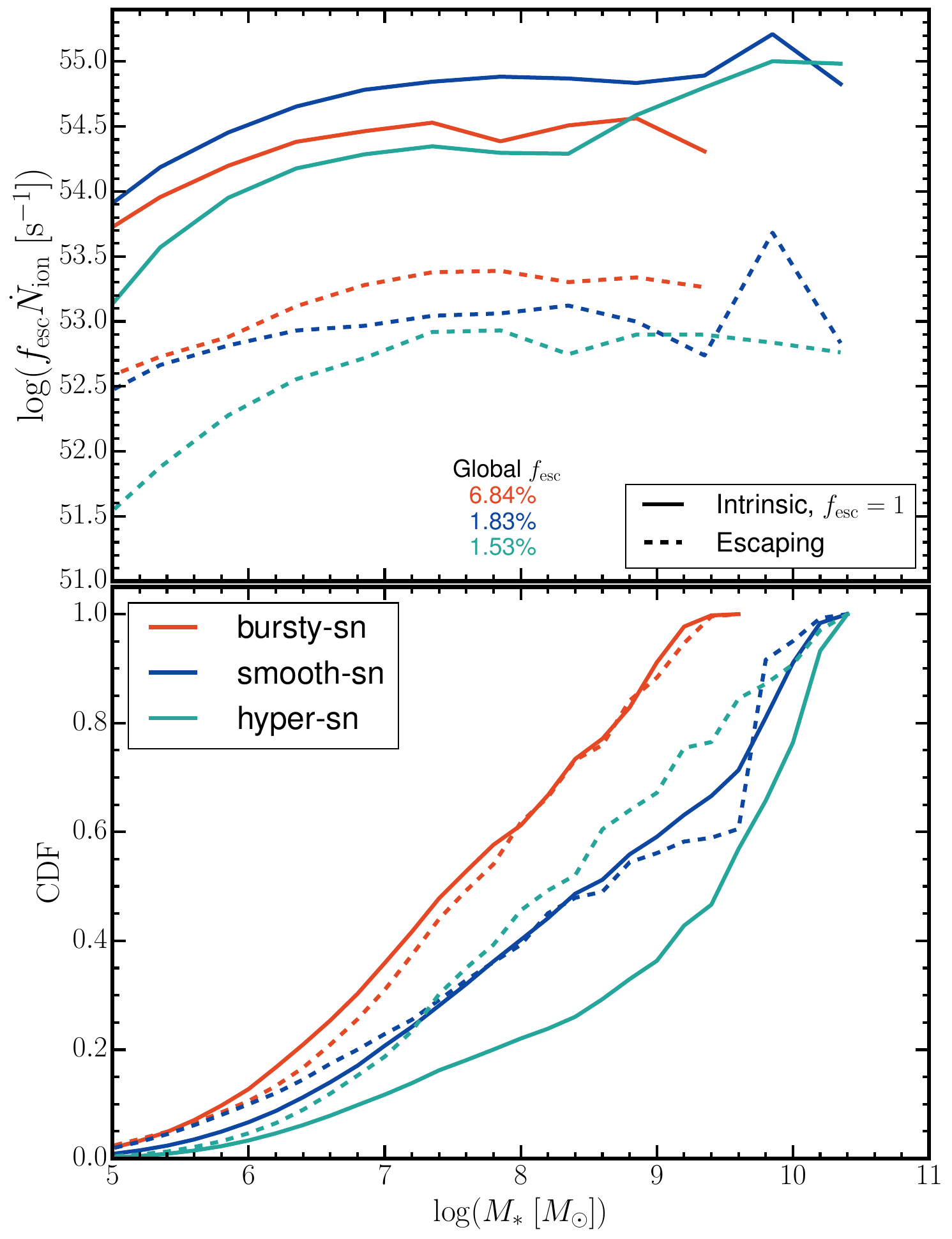}
    \caption{\textbf{Top}: Intrinsic (solid lines) and escaping (dashed line) ionising emissivity as a function of stellar mass for {\tt bursty-sn} (red), {\tt smooth-sn} (blue) and {\tt hyper-sn} (turquoise). \textbf{Bottom}: Cumulative distribution of the above quantities indicating net contributions from galaxies in different mass bins. Emissivities are calculated over the full duration of the simulation.}
    \label{fig:NionXiion}
\end{figure}

We next quantify differences in reionisation histories by looking at the net intrinsic and escaping emissivities from all galaxies, defined as $f_{\rm esc} \dot{N}_{\rm ion}$, where $\dot{N}_{\rm ion}$ is the production rate of photons with $E_\gamma>13.6$eV calculated by integrating the stellar SED, and $f_{\rm esc}=1$ for the instrinsic emissivity.
The top panel of Figure~\ref{fig:NionXiion} shows the total intrinsic (solid lines) and escaping (dashed lines) emissivities of the galaxies over the entire duration of the simulation as a function of stellar mass. As previously discussed, the net intrinsic emissivity is the highest for {\tt smooth-sn}, followed by {\tt hyper-sn} and {\tt bursty-sn}. The intrinsic contributions for {\tt smooth-sn} and {\tt hyper-sn} are largely dominated by the most massive galaxies (see also discussion below), whereas {\tt bursty-sn} shows a more even distribution across the mass range. However, the net escaping emissivity is the highest for {\tt bursty-sn}, followed by {\tt smooth-sn} and finally {\tt hyper-sn}. We can also use the total intrinsic and escaping emissivity to calculate a global escape fraction for each simulation, finding $\approx 6.8, 1.8$ and $1.5\%$ for {\tt bursty-sn}, {\tt smooth-sn} and {\tt hyper-sn}, respectively. 

To understand which galaxies contribute the largest to reionisation, we look at the cumulative distribution of emissivities, which are shown in the bottom panel of Figure~\ref{fig:NionXiion} as intrinsic (solid lines) and escaping (dashed lines).  {\tt bursty-sn} results in a fairly even contribution from galaxies with $M_* > 10^7 \rm M_\odot$ to both intrinsic and escaping emissivities. Differently, the intrinsic emissivity in {\tt smooth-sn} and {\tt hyper-sn} is largely dominated by galaxies with $M_* \gtrapprox 10^9 \rm M_\odot$ ($\approx 50\%$ and $ \approx 65\%$, respectively). While the escaping emissivity shows a distribution similar to the intrinsic one (i.e. dominated by galaxies with $M_* \gtrapprox 10^9 \rm M_\odot$), {\tt hyper-sn} exhibits an even contribution from galaxies with $M_* > 10^7 \rm M_\odot$. Therefore, in {\tt bursty-sn} and {\tt hyper-sn}, galaxies with $M_* > 10^7 \rm M_\odot$ contribute evenly to reionisation, whereas in {\tt smooth-sn} $\approx 46\%$ of the escaping emissivity is produced by $M_* > 10^9 \rm M_\odot$ galaxies. 
 
\begin{figure}
    \includegraphics[width=\linewidth]{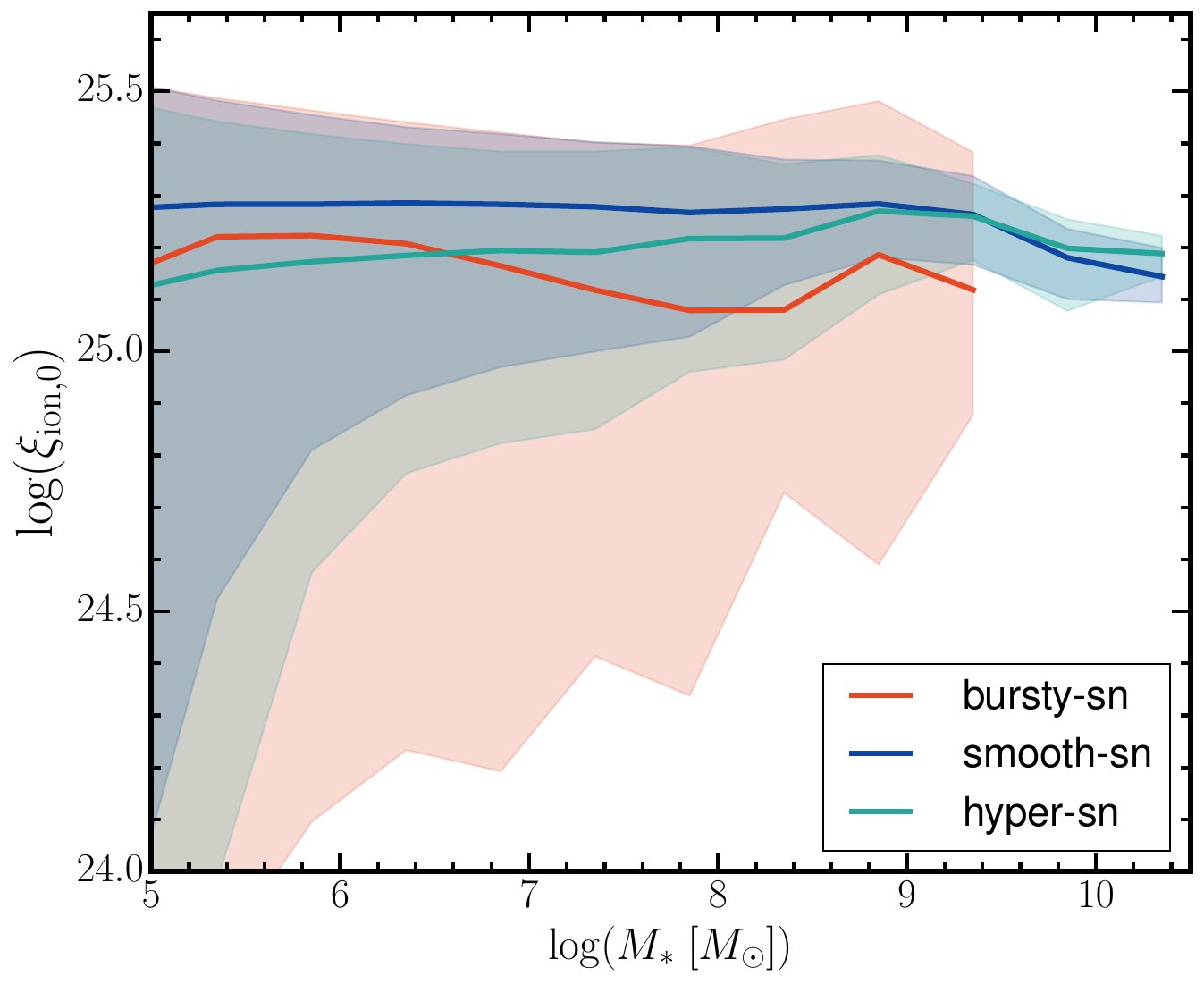}
    \caption{The median ionising photon production efficiency $\xi_{\rm ion,0}$ as a function of stellar mass. Scatter around the median represents 16th and 84th percentiles. $\xi_{\rm ion,0}$ is calculated over the full duration of the simulations. All models show very similar $\xi_{\rm ion,0}$, although leading to very different reionisation histories suggesting that reionisation is $f_{\rm esc}$ limited.}
    \label{fig:Epsion}
\end{figure}

As measuring $\dot{N}_{\rm ion}$ at high redshifts is challenging, other quantities  are used to evaluate the ionising photon production, such as the ionising photon production efficiency $\xi_{\rm ion,0}$ = $\dot{N}_{\rm ion}/L_{1500}$, where  $L_{1500}$ is the intrinsic luminosity of the galaxy at 1500~\AA. $\xi_{\rm ion,0}$ can give key insights about the stellar populations and dust obscuration and is used to study escape fractions of high redshift galaxies \citep{Simmonds_2023}. In Figure~\ref{fig:Epsion}, we show $\xi_{\rm ion,0}$ as a function of stellar mass of galaxies calculated over the full duration of the simulation. All three models show comparable median trends with minor differences of $\approx (0.1-0.3)$ dex, however, {\tt bursty-sn} has a very large scatter at all masses. Previous estimates from the HST surveys \citep{Robertson_2013} and more recent JWST programs \citep{atek2023jwst,Simmonds_2023,simmonds2023lowmass,endsley2023starforming,Endsley_2023,saxena2023jades} find median values of log($\xi_{\rm ion,0}$) $\approx24.8-25.7$ for galaxies in the range $z\sim5-11$, which is in agreement with all three models. 

Despite a very similar ionising photon production efficiency, the three models result in quite different reionisation histories (see section~\ref{sec:reionhist}), suggesting that reionisation is strongly $f_{\rm esc}$ limited. Indeed, we have seen that different feedback models produce a different morphological mix (bottom panel in Figure~\ref{fig:VSigma}), which in turn affects radiation escape depending on the relative predominance of rotation- or dispersion-dominated systems, and hence the escaping ionising emissivity budget. Therefore, models such as {\tt smooth-sn} which preferentially produce rotationally-dominated systems (especially the most massive and luminous galaxies) show very low global escape fractions and produce very late reionisation histories. Meanwhile, {\tt bursty-sn} produces a majority of dispersion-dominated systems (especially the most massive and luminous galaxies) which implies a high global $f_{\rm esc}$ ($\approx 4\times$ higher than {\tt smooth-sn}) and reionises the simulation volume by $z=5.1$.

\section{Discussion}
\label{sec:discussion}
Here we discuss the implications of feedback variations on galaxy properties, the prospects of constraining stellar feedback using observations, and the limitations of our work. 
\subsection{Connecting galaxy morphology and LyC escape: insights for observations}

We have shown that the SN feedback mode hardly affects some observables (e.g SFMS, UVLF), while strongly altering others (e.g SFRD, SFE, reionisation, morphologies, $f_{\rm esc}$). A strong indicator to test feedback models we point to is the morphological mix that emerges post-reionisation, as feedback alters in a fundamental way stellar and gas morphologies. The gas morphologies in turn affect the $f_{\rm esc}$ of galaxies. As galactic morphology can be probed  using different tracers (including stellar light and emission line kinematics), a multi-wavelength study of morphological characteristics provides a key insight to constrain feedback models (see Figure~\ref{fig:reion-proj}). 

Telescopes like JWST and ALMA allow us to map galaxies at high angular resolutions ($\approx0.1-0.2$ arcsec) deep into the epoch of reionisation. Studies using JWST observations have already started to produce galactic morphology statistics at various redshifts using e.g. stellar surface brightness profiles, Sersic indices, and Gini indices  
\citep{huertascompany2023galaxy, Treu_2023, 2023Jacobs, Kartaltepe_2023, vegaferrero2023nature,Tacchella_2023,sun2023structure}.  
All find a rich morphological diversity of galaxies well established already at $z>5$, and estimate a disk (pure disk + disks with bulges) fraction in the range $\sim30-40\%$ at $z\sim6$, in agreement with our {\tt busty-sn} model. 
In this work we just consider dispersion- or rotation-dominated galaxies, finding that their relative distribution varies strongly depending on the feedback model. We defer to a future work a more direct comparison to observations, which will help to better constrain our models. 

Recent studies have used SED fitting \citep{looser2023jades,dressler2023building}, \textsc{[OIII]+H$\beta$} equivalent width and H$\alpha$ emission line luminosities \citep{endsley2023starforming,Endsley_2023, simmonds2023lowmass} to characterise modes of star formation in JWST galaxies, finding that a bursty SFH is prevalent in galaxies with masses $M_* \sim 10^{7-10} \rm M_\odot$ in the range $z=6-12$. These studies also show that galaxies with a bursty SFH are likely to drive reionisation. In sections~\ref{sec:reionhist} and \ref{sec:morphologies} we have discussed the strong influence that a bursty SFH has on galaxy morphology and on the reionisation history, supporting the claim that galaxies with bursty SFH are the likely drivers of reionisation. While observations have not yet connected galaxy morphologies to their ionising efficiencies, in {\tt SPICE} we find that galaxies with a bursty SFH are largely dispersion-dominated (especially evident at $z=5$), while galaxies with a non-bursty or smooth SFH are mainly rotation-supported and are unlikely to drive reionisation (see section~\ref{sec:Lycfesc}). Such prediction can be tested with JWST observations. 

The impact of cold streams \citep{Dekel_2009,Bournaud_2009,Oh_2018}, gas fractions \citep{1996Barnes,2001Barnes,Naab_2006}, mergers \citep{1977Toomre,Hopkins_2009,Kannan_2015,Flores_Vel_zquez_2020}, tidal effects \citep{1998Bekki} and stellar feedback \citep{Okamoto_2005,Agertz_2016} on galaxy morphologies has been a topic of debate over the last few decades. While it is widely accepted that disk galaxies form due to gas accretion and transform to spheroidal galaxies via mergers \citep{Toomre1972,1977Toomre,Naab_2006}, the picture becomes unclear at high redshifts. For instance, the short dynamical timescales at high redshifts along with the long depletion timescales can allow for disk galaxies to stabilize even after major mergers \citep{2006Robertson}.  
As in all {\tt SPICE} simulations the initial conditions are the same, we expect the timing of halo mergers as well as the cold streams feeding the  haloes to be very similar across all models. Therefore, we argue that the stark differences in galaxy properties are a consequence of supernova feedback systematically altering the structure of galaxies on ISM/CGM scales. These differences translate not only to galaxy morphology (see section~\ref{sec:morphologies}), but also to the LyC escape fractions (see section~\ref{sec:Lycfesc}). 

As a direct measurement of LyC escape fractions from high redshift galaxies is not possible because of the high IGM opacity, alternative indirect diagnostics have been suggested, such as metal line ratios  \citep{2019WangSiII,Chisholm_2020,2022SaxenaCIV,2022SchaererLyC}, Ly$\alpha$ peak separations \citep{Verhamme_2015,2017Verhamme} and recent SFR \citep{calzetti2012star,Kennicutt_2012,Flores_Vel_zquez_2020}. While all these probes are sensitive to dust attenuation, line strengths and orientation effects, galaxy morphology can be reliably traced with high resolution observations, and could thus be used as a better proxy for the escape fraction. Indeed, here we show that a strong correlation exists between the morphology  of galaxies (described in terms of $V/\sigma$) and their escaped radiation, with dispersion-dominated systems exhibiting the highest LyC escape fractions at all redshifts (by factor of 10-50). Our results  also indicate that, as reionisation is limited by the escape fraction, it is strongly dependent on the behavior of stellar feedback, as this needs to be explosive and bursty (see section~\ref{sec:reionhist},~\ref{sec:Lycfesc}). 

Due to the differences in LyC $f_{\rm esc}$, at the same intrinsic luminosity, dispersion-dominated galaxies should preferentially create large expanding HII regions as compared to their rotation-dominated counterparts (see Figure~\ref{fig:reion-proj}). The growth of the HII regions is thus connected to the galaxy morphology, suggesting that concurrent observations of galaxy morphology and ionised regions could help to establish more firmly the connection between morphology and $f_{\rm esc}$, and  to constrain stellar feedback. Observations of Ly$\alpha$ emitters (LAE) from HST/Keck-MOSFIRE \citep{Stark_2016,MattheeCOLA1} suggest that bright LAEs are preferentially surrounded by large ionised regions, and also characterised by intense star formation. However, recent JWST observations of LAEs at $z\sim7-9$ \citep{endsley2023starforming,whitler2023insight,tang2023jwstnirspec,Intae2023,Saxena_2023bubble} suggest a more diverse HII regions distribution, where not all strong LAEs lie within the largest ionised regions. Therefore, deep spectroscopic observations combined with imaging of LAEs will help to better constrain bubble sizes along with galactic properties. In a companion study (Bhagwat et al. in prep.) we will investigate Ly$\alpha$ characteristics of galaxies in the three feedback models. 

Understanding the demographics of outflows driven by SN feedback and radiation pressure \citep{Hayward_2016,Li_2017,Costa2018b,Menon_2023} is key to constrain feedback models at high redshifts. Recent JWST observations \citep{zhang2023statistics, carniani2023jades} investigate the incidence of ionised outflows using H$\alpha$ and {\tt [O III]} in low mass galaxies ($M_* < 10^{10} \rm M_\odot$) in the range $z\sim6-9$. These studies find that the inferred outflow velocities and mass loading factors are, respectively, 3 and 100 times larger than those observed for local dwarfs \citep{Marasco_2023}.  Multi-wavelength investigations combining JWST and ALMA data ({\tt [O III]} and {\tt [C II]}, respectively; e.g. \citealt{fujimoto2022jwst}) suggest that the incidence of an outflow from a $M_* \approx 10^{8} \rm M_\odot$ galaxy at $z\sim8.5$ is likely associated to a starburst which consequently promotes strong ionising photon escape. It is important to note, however, that in obscured systems (such as SMG/ULIRGs) ionising photon escape into the IGM can be strongly suppressed due to optical depth effects despite incidence of outflows \citep{Cen_2020}. Moreover, non-bursty (but high) star-formation (as seen in the {\tt smooth-sn} model) fails to drive strong outflows, which allows the gas to settle within the disk boosting the gas fractions of galaxies and consequently suppressing $f_{\rm esc}$ (see \citealt{2020Yoo} for similar findings).`
In {\tt SPICE} we have shown how feedback affects the outflow characteristics (see Figure~\ref{fig:Tmetz5}) suggesting that outflow properties and incidence rates can be a key tool to constrain feedback models. 

\subsection{Numerical uncertainties and physical limitations}
\label{sec:numerics}

The range of physics included in {\tt SPICE}, along with a physical resolution of $\sim 28$ pc, allows to make robust predictions for a variety of galaxy observables. However, due to approximations in numerical schemes, subgrid models and stochasticity, {\tt SPICE} is affected by uncertainties. Understanding them is critical to improve models for future simulations, as well as the interpretation of their predictions. 

While {\tt SPICE} models metal line cooling, we do not include a molecular line cooling channel, which is relevant in the dense gas phase of the ISM. This can potentially impact, among others, the presence of cold gas in the outflows driven by SN feedback \citep{Richings2014I,Richings2014II,Biernacki_2018}, as well as the distribution and efficiency of star formation. 
While the refinement strategy adopted in {\tt SPICE} allows us to resolve the CGM of galaxies at $\sim(28-84) \rm pc$ scales, the resolution deteriorates further away from the centers of haloes, with possible consequences on the correct modeling of the energy, mass loading and multiphase nature of outflows \citep{rey2023boosting}. This in turn can have important effects on the modeling of the escape of LyC radiation.  

We attempted to improve the feedback modelling by making it increasingly physically based. {\tt bursty-sn} represents the "IMF averaged" model, whereas {\tt hyper-sn} includes realistic SN energies and explosion times, along with a theoretically supported prescription for hypernovae \citep{sukhbold2016core,kobayashi2006galactic,grimmett2020chemical}. However, we note that the latter, which is the most physically motivated model, is unable to complete reionisation by the end of the simulation, whereas {\tt bursty-sn} (the least physically motivated model) is more consistent with observational constraints. While this could point to issues in the assumed HN rates or SN energy distributions, it could also indicate a potentially missing ingredient in the feedback modelling (e.g. runaway stars \citealt{Andersson2020,Steinwandel2023}), or the need for an improved numerical scheme for injection of energy and momentum.

The ISM model used in {\tt SPICE} is a combination of subgrid prescriptions based on idealised simulations 
(e.g turbulence from \citealt{Federrath_2016} and mechanical SN feedback from \citealt{kimm2014escape}). These prescriptions include a number of correction terms that are not predicted ab initio, but rather derived from smaller-scale simulations. Therefore, central assumptions such as log-normal density PDF and missing feedback processes at sub-resolution scale are a source of uncertainties. 

Detailed studies of population synthesis models and SEDs have shown that the SED choice can strongly affect LyC escape fractions and reionisation histories \citep{Ma_2016,SphinxRosdahl2018,2022Qingbo,rosdahl2022}. Further observational evidence \citep{Gotberg2019,Gotberg2020,Gotberg2023StrippedB} suggests that e.g. the contribution from stripped binary stars could boost the ionising photon budget from massive stars. Hence, the SED and the ingredients of populations synthesis models are important uncertainties which should be further explored in future work. 

{\tt SPICE} employs the reduced speed of light approximation to advect radiation. This approximation produces converged reionisation histories as long as the speed of the ionisation fronts (I-fronts) remains below the value of the adopted reduced speed of light, which in {\tt SPICE} is $0.1c$. We note that at the tail end of reionisation, the I-fronts can travel as fast as $0.1 c$ as they traverse the voids \citep{2019DAloisio}, but the approximation is unavoidable because of the extreme computational costs of running simulations with the full speed of light. 

We also note that the simulation volume is too small to study the global reionisation on large scales. Indeed, typical sizes of ionised regions can become tens of cMpc, suggesting that to produce converged reionisation histories boxes $\gtrsim 100$cMpc/$h$ are required  \citep{Iliev_2014,gnedin2022modeling}. In follow-up studies we plan on extending the {\tt SPICE} models to larger simulation volumes to get better galaxy statistics and to further improve numerical schemes of stellar feedback.  

{\tt SPICE} addresses the effect of variations in stellar feedback models while including some poorly-explored physical processes, like radiation pressure on dust and dust itself. However, due to the computational costs involved, our models are far from complete, as processes such as cosmic rays (CRs), AGN feedback and magnetic fields are still missing. Magnetic fields and CRs can both affect the structure of the ISM which in turn affects galaxy obervables. Numerous studies have attempted to quantify the effect of magnetic fields \citep{Gnedin_2000mag,Marinacci_2015,krumholz2019role,McKee_2020,Garaldi2021Magnetic,Katz_2021,Martin_Alvarez_2020} and CRs \citep{Sazonov_2015,Leite_2017,farcy2022,alvarez2022} on galaxy formation and reionisation itself. However, the implications remain unclear. 

Dust is relevant for various processes such as cooling, fragmentation, absorption and scattering of radiation and obscuration of UV photons.
{\tt SPICE} adopts a simplified dust model in which dust is hydrodynamically coupled to the gas and its creation and destruction are not accounted for. Additionally,  we assume that dust opacities are independent of temperature, although in dense cold gas,
they can scale as $T^2$ \citep{Semenov2003}. Scaling of opacities can also strongly affect the radiation pressure on dust, which we include as a feedback channel. Therefore, on-the-fly modelling of dust grain creation and destruction, as well as  more accurate  dust opacities, are important for a better modelling of the ISM structure and of the strength of radiation pressure driven feedback.

Theoretical studies have argued that AGN feedback can be important to understand the evolution of dwarf galaxies, even at very high redshifts \citep{Dashyan_2017,Koudmani_2021,Koudmani_2022, Sharma2022}. At the same time, observational evidence is mounting supporting the presence of a population of $\sim (10^5 - 10^8) \ \rm M_\odot$ black holes at very high redshifts \citep{scholtz2023gnz11,Mezuca2023, 2023Schneider, maiolino2023jades}. Therefore, AGN feedback is a key missing ingredient worth exploring in the future to investigate when and in which galaxies it is relevant.  
\section{Conclusions}
\label{sec:conclusion}
In this work we introduce {\tt SPICE}, a suite of radiation hydrodynamical simulations of galaxy formation and reionisation performed with {\tt RAMSES-RT}. Our aim is to systematically study the effects of variations in stellar feedback on reionisation and properties of the first galaxies. {\tt SPICE} resolves atomic cooling haloes and has a spatial resolution of $\approx28$ pc at $z=5$. The galaxy formation model  includes cooling, non-equilibrium chemistry, a multi-freefall star formation model, which employs a subgrid prescription for turbulence to evaluate the star formation efficiency in each computational cell, and a mechanical feedback scheme to inject energy and mass from supernovae. We include radiation transport (through gas and dust) in 5 frequency bands, i.e. IR, optical and three UV groups. 
Our base model remains identical across our different simulations with the sole exception of SN feedback, which is modelled in three different way (see Table~\ref{tab:simulations}): bursty ({\tt bursty-sn}), smooth ({\tt smooth-sn}) and smooth feedback with a time-evolving energetic component ({\tt hyper-sn}). 

In this introductory paper, we showcase the global properties of the simulations in terms of galaxy populations, reionisation histories, and connection between galaxy morphologies and LyC escape fractions. A summary of our key findings is presented below:

\begin{itemize}    
    \item Feedback strongly affects the burstiness and amplitude of the star formation rate density (SFRD; section~\ref{sec:starformation}). {\tt bursty-sn} consistently exhibits bursts of SF, with a more pronounced intensity below $z=8$. Conversely, {\tt smooth-sn} has a minimal burstiness, consistently reaching the highest SFRD among the three models. Finally, in {\tt hyper-sn}, the SFRD is the lowest at the highest redshifts because of the strong feedback from hypernovae. This also induces pronounced burstiness at $z>9$, akin to {\tt bursty-sn}. As the hypernova rate declines (due to metal enrichment), below $z\sim9$ the model exhibits similarities with {\tt smooth-sn} and shows a strong upswing in SFRD.
    
    \item We observe the emergence of a star formation main sequence (SFMS; section~\ref{sec:starformation}) already by $z=10$. The median SFMS remains very similar across models, however, {\tt bursty-sn} exhibits a much larger scatter ($0.3-0.5$ dex higher) as compared to the other two models. All three models show good agreement with observations from various JWST programs. 
    
    \item Feedback strongly affects reionisation histories (section~\ref{sec:reionhist}).  {\tt bursty-sn} completes reionisation by $z=5.1$, consistent with observational constraints. This is not the case for {\tt smooth-sn} and {\tt hyper-sn}, despite both models yielding an excess of ionising photons in comparison to {\tt bursty-sn}. We thus confirm that reionisation is very sensitive to the modulation of $f_{\rm esc}$ due to feedback, rather than being driven solely by the number of available ionising photons. 
    
    \item The impact of feedback is visible also in the intrinsic $1500$~\AA~luminosity function (LF; section~\ref{sec:luminosityfunc}). While in {\tt smooth-sn} and {\tt hyper-sn} we observe an abundance of highly UV-bright galaxies ($M_{1500} < -23$), these are absent in {\tt bursty-sn}. The {\tt hyper-sn} ({\tt smooth-sn}) model consistently yields the lowest (highest) galaxy counts at all redshifts. 
    
    \item The evolution of the dust-attenuated LFs is different from that of the intrinsic ones (section~\ref{sec:luminosityfunc}). Interestingly, while only in {\tt smooth-sn} we observe signs of dust attenuation at $z=10$, by $z=7$ all models exhibit noticeable levels of attenuation, which becomes even more pronounced at $z=5$. All three models show an excellent agreement with observations, especially at $z=7$ and 5. Despite intrinsic LFs being different, dust-attenuated ones show minimal discrepancies  across models. Therefore, LFs cannot be directly used to constrain feedback models. 
    
    \item We see striking differences in post-reionisation galaxy morphologies, characterized by the ratio $V/\sigma$ (section~\ref{sec:morphologies}), where we classify galaxies with $V/\sigma \geq 1$ as rotationally-supported and dispersion-supported otherwise. {\tt bursty-sn} exhibits a preference for dispersion-dominated systems, primarily at the higher mass end, while in {\tt smooth-sn} most galaxies are rotationally-dominated at any mass. Finally, in {\tt hyper-sn} galaxies with masses below (above) $M_* \approx 10^8 \rm M_\odot$ typically show strong dispersion (rotational) support.
    
    \item The differences in morphology translate into variations in  Lyman continuum (LyC) escape fractions ($f_{\rm esc}$; section~\ref{sec:Lycfesc} and Figure~\ref{fig:fescvsMstar}). We find that $f_{\rm esc}$ of dispersion-dominated systems is enhanced by a factor of $\approx10-50$ in comparison to that of rotation-dominated ones at all redshifts. 
    
    \item Galaxies that undergo strong bursts of feedback preferably produce dispersion-dominated systems (especially at the highest mass and luminosities), and hence higher $f_{\rm esc}$ as compared to galaxies that have smoother feedback, which allow for the formation of stable, rotationally-supported systems (in particular at the highest mass and luminosities). Therefore, feedback determines the global morphological mix of galaxies, and, as a consequence, the global $f_{\rm esc}$. The connection between morphologies and $f_{\rm esc}$, therefore, can potentially be an excellent probe not only of feedback models at high redshifts, but also of the galaxies that drive reionisation. 

    \item The intrinsic contribution to ionising emissivity ($\dot{N}_{\rm ion}$; solid lines in Figure~\ref{fig:NionXiion}) is dominated by galaxies with $M_* \gtrsim 10^9 \rm M_\odot$ in {\tt hyper-sn} ($\approx65\%$) and {\tt smooth-sn} ($\approx50\%$), whereas {\tt bursty-sn} shows a similar contribution from galaxies of all stellar masses above $\sim 10^7 \rm M_\odot$. 
    
    \item The escaping emissivity ($f_{\rm esc}\dot{N}_{\rm ion}$, dashed lines Figure~\ref{fig:NionXiion}) in the three models is dominated by galaxies in different mass ranges. In {\tt smooth-sn} the largest contribution ($\approx 46\%$) comes from  $M_* > 10^9 \rm M_\odot$ galaxies, whereas {\tt bursty-sn} and {\tt hyper-sn} show roughly uniform contribution from galaxies in the range $M_* \approx 10^{7-10} \rm M_\odot$. Upon comparing the escaping and intrinsic emissivities, we find a global (i.e. from the entire galaxy populations at all redshifts) escape fraction of $\approx 6.8, 1.8$ and $1.5 \%$ for {\tt bursty-sn}, {\tt smooth-sn} and {\tt hyper-sn}, respectively. 

    \item  The ionising photon production efficiency ($\xi_{\rm ion, 0} = \dot{N}_{\rm ion}/ L_{\rm 1500}$) evaluated over the full duration of the simulation shows comparable median trends (see Figure~\ref{fig:Epsion}) for the three models, with differences of $\approx 0.1-0.3$ dex. Despite a very similar ionising photon production efficiency, the three models result in quite different reionisation histories, suggesting that reionisation is strongly $f_{\rm esc}$ limited. 
\end{itemize}

The feedback variations as well as the high resolution of {\tt SPICE} allow us to probe in detail the first sources of reionisation in their stellar, gas and radiative components. In this work, we focused on the morphology and LyC $f_{\rm esc}$ to connect feedback to reionisation using kinematics derived from stars and H$\alpha$ bright gas. However, imprints of feedback will manifest in a wide range of observables. The multi-phase ISM model, along with the additional radiation transport in IR and optical, allows us to investigate a variety of emission lines (such as {\tt H$\alpha$}, {\tt [C II]}, {\tt [O III]} and Ly$\alpha$).  
Indeed, all these probes  offer a unique opportunity to make predictions for and comparisons to observations across a wide range of wavelengths, from state-of-the-art facilities such as JWST, ALMA and MUSE, as well as from planned ones such as SKA and ELT. Therefore, a comparison between synthetic {\tt SPICE} observables and real data will help to disentangle and understand different feedback mechanisms at high redshifts.  

\section*{Acknowledgements}
The authors thank the anonymous referee for an insightful report and constructive comments that helped improve this work. Computations were performed on the HPC systems Raven and Cobra at the Max Planck Computing and Data Facility (MPCDF). AB would like to thank the MPCDF team for their support with HPC tickets.  The authors would like to extend a great deal of gratitude to Joakim Rosdahl for helping with {\tt RAMSES-RT} setups and sharing analysis pipelines employed in this work. The authors are grateful to Taysun Kimm for developing and sharing the mechanical feedback model adopted and modified in this work. AB would also like to thank Christian Partmann, Eileen Herwig, Volker Springel, Chiaki Kobayashi, Thorsten Naab, Martyna Chruślińska, Ylva G$\Ddot{\rm o}$tberg, Thomas Janka and Selma de Mink for helpful discussions; Jeremy Blaizot and Leo-Michel Dasnac for developing {\tt RASCAS};  Romain Teyssier for developing the code {\tt RAMSES} and making it public. This work made extensive use of multiple publicly available software packages:  {\tt matplotlib} \citep{hunter2007matplotlib}, {\tt numpy} \citep{van2011numpy}, {\tt scipy} \citep{scipy} and {\tt CoReCon} \citep{2023CoreCon}. The authors would like to thank the community of developers and those maintaining these packages.

\section*{Data Availability}

Data will be provided on reasonable request to the authors.



\bibliographystyle{mnras}
\bibliography{biblio.bib} 




\appendix

\bsp	
\label{lastpage}
\end{document}